\documentclass[10pt,journal,  oneside, twocolumn]{IEEEtran}

\usepackage{multirow}
\usepackage{latexsym}
\usepackage{graphicx}
\usepackage{float}
\usepackage{amsmath}
\usepackage{amsthm}
\usepackage{lipsum}
\usepackage{subcaption}
\usepackage{graphicx}
\usepackage{authblk}
\usepackage{bm}
\usepackage{booktabs}
\usepackage{amsthm}
\usepackage[section]{placeins}
\usepackage{soul}
\usepackage{tabularx}

\usepackage{colortbl}

\usepackage{enumitem}

\usepackage{CJKutf8}

\usepackage[numbers,sort&compress]{natbib}

\makeatletter  
\newif\if@restonecol  
\makeatother

\usepackage[linesnumbered,ruled,vlined]{algorithm2e}
\usepackage{algpseudocode}  
\usepackage{amsmath}

\usepackage{amssymb}

\usepackage{mathrsfs}
\usepackage{subfig}
\usepackage{caption}
\newcounter{problem}

\renewcommand{\baselinestretch}{1.0}

\begin{document}

\title{Information-Optimal Formation Geometry Design for Multimodal UAV Cooperative Perception}

\author{Kai Xiong, Xingyu Wu, Anna Duan, Gang Li, Yongjun Huang, Supeng Leng,~\IEEEmembership{Senior Member,~IEEE}, Jianhua He,~\IEEEmembership{Senior Member,~IEEE}


\thanks{

K. Xiong, X. Wu, G. Li, and S. Leng are with School of Information and Communication Engineering, University of Electronic Science and Technology of China, Chengdu, 611731, China.
}

\thanks{
A. Duan is with AVIC Chengdu Aircraft Design \& Research Institute, 610041, Chengdu, 610041, China.
}

\thanks{
J. He is with School of Computer Science and Electronic Engineering, University of Essex, Colchester, UK.
}

\thanks{The financial support of Sichuan Provincial Natural Science Foundation under Grant 2026NSFSC1431, National Natural Science Foundation of China (NSFC), Grant No.62201122, and AVIC United Technology Center for Intelligent Decision-making and Coordinated Control Mechanism Model Research grant.

}

\thanks{The corresponding author is Supeng Leng, email: spleng@uestc.edu.cn}
}


\maketitle

\begin{abstract}
The efficacy of unmanned aerial vehicle (UAV) swarm cooperative perception fundamentally depends on three-dimensional (3D) formation geometry, which governs target observability and sensor complementarity. 
In the literature, the exploitation of formation geometry and its impact on UAV sensing have rarely been studied, which can significantly degrade multimodal cooperative perception in scenarios where heterogeneous payloads (vision cameras and LiDAR) should be geometrically arranged to exploit their complementary strengths while managing communication interference and hardware budgets. 
\textcolor{black}{To bridge this critical gap, targeting the active tracking phase where a prior target estimate is available, we propose an information-optimal optimization framework that optimizes the allocation of UAVs and multimodal sensors, configures formation geometries, and executes flight control.}
The UAV-sensor allocation is optimized by the Fisher Information Matrix (FIM) determinant maximization.
Under this framework we introduce an equivalent formation transition strategy that enhances field-of-view (FOV) coverage and reduces dynamic communication interference without compromising perception accuracy. 
Furthermore, we design a novel Lyapunov-stable flight control scheme with logarithmic potential fields to generate energy-efficient trajectories for formation transitions. Extensive simulations demonstrate that our formation-aware design achieves 25.0\% improvement in FOV coverage, 104.2\% enhancement in communication signal strength, and 84.7\% reduction in energy consumption compared to conventional benchmarks. A high-fidelity Gazebo validation further shows up to 20.2\% gain in perception accuracy over hemispherically sampled formations. \textcolor{black}{These results suggest that task-driven geometric allocation is an important rather than incidental component in next-generation UAV swarm perception systems.}

\end{abstract}


\IEEEpeerreviewmaketitle

\section{Introduction}
\IEEEPARstart {U}{n}manned Aerial Vehicle (UAV) swarms have emerged as transformative platforms for cooperative perception.
They offer unprecedented capabilities in cooperative target tracking, surveillance, and the active-sensing phase of search and rescue (SAR) operations where a prior target state estimate is available \cite{8660516,7941945,10557126}.
Unlike single-UAV systems or terrestrial sensor networks, aerial swarms provide enhanced spatial coverage, improved viewpoint diversity, and parallel multimodal data acquisition through collaborative sensing \cite{8784232}.
It is noted to distinguish between two different operational paradigms in SAR tasks: the \emph{blind search} paradigm, where no prior information about the target exists, and the \emph{active tracking} paradigm, where a preliminary target state estimate has been obtained through initial detection (e.g., distress signals, thermal signatures, or prior sweeps) \cite{zhaoOptimalSensorPlacement2013}. 
This paper focuses on the latter one, in which the UAV swarm has access to a predicted target state and seeks to optimize its formation geometry for the information gain of subsequent observations. \textcolor{black}{Here, the active tracking paradigm is consistent with the predictor-corrector framework widely adopted in information-driven sensor management \cite{6018982}. In this framework, formation planning is conditioned on the best available state estimate, rather than requiring exact target knowledge.}

However, current active tracking literature concentrates on trajectory planning, resource allocation, and perception algorithms while treating the formation geometry of UAV swarms as a secondary implementation detail \cite{{10492856},{10556867}, {10628042},{10089158}}. 
\textcolor{black}{This oversight has significant impact on multimodal cooperative perception performance in scenarios where heterogeneous payloads (vision cameras and LiDAR) are arranged to exploit their complementary strengths. 
Specifically, cameras provide rich visual semantics but suffer from depth estimation limitations, while LiDAR delivers precise three-dimensional (3D) geometric information at the cost of higher communication overhead and energy consumption \cite{LEE2013105}. 
Accordingly, the formation geometry can govern field-of-view (FOV) coverage, heterogeneous sensor complementarity, and target observability.
Without proper geometric allocations, these complementary sensing capabilities cannot be effectively leveraged, leading to redundant observations and coverage blinds.}

{\color{black}Moreover, the geometric formation optimization is also challenged by practical collaboration constraints.
In SAR tasks, UAVs typically operate in proximity, creating significant communication interference in dense formations \cite{9385994}. 
While increasing swarm size enhances perception quality, it can significantly raise communication load, computational burden, and hardware costs.
This requires a fundamental trade-off between multimodal perception performance and resource efficiency. It cannot be resolved through trajectory optimization alone.
Furthermore, existing formation strategies typically assume homogeneous UAV capabilities \cite{panwarOptimalSensorPlacement2022a, zhaoOptimalSensorPlacement2013}, without considering the spatial arrangement requirements of heterogeneous sensor payloads, which is a critical limitation for multimodal perception systems.

\textcolor{black}{To bridge this critical gap, we propose an information-optimal framework that jointly optimizes geometric allocation and sensor heterogeneity through Fisher Information Matrix (FIM) determinant maximization. 
The proposed formation optimization is activated after an initial target estimate is obtained.
By treating formation design as a primary optimization variable rather than a byproduct of trajectory planning, we achieve simultaneous improvements in coverage completeness, communication Signal-to-Interference-plus-Noise Ratio (SINR), and flight energy efficiency.}
The main contributions of this paper are as follows:
}

\begin{itemize}

\item We establish a novel information-optimal framework for 3D UAV swarm formation design that explicitly quantifies the interdependence between spatial geometry, heterogeneous sensor modalities, and cooperative perception performance. By leveraging FIM determinant maximization as the optimization criterion, our framework provides a principled approach to determine information-optimal formation allocations in versatile perception environments where conventional homogeneous formation strategies fail to exploit multimodal sensing synergies.

\item We formulate and solve a formation transition problem that addresses the FOV-oriented formation allocation under practical operational constraints.
The proposed optimization model uniquely balances sensing information gain against FOV coverage completeness, communication overhead, and sensor payload costs. It incorporates realistic constraints including directional FOV limitations, inter-UAV interference effects, and heterogeneous payload capabilities that are neglected in existing literature.

\item We develop a Lyapunov-stable distributed flight controller for energy-efficient formation transitions during dynamic operations. The control scheme employs a logarithmic potential field to ensure smooth trajectories with bounded control authority. This design provably minimizes unnecessary maneuvers and energy consumption while guaranteeing asymptotic flight stability. Extensive simulations show that our integrated approach achieves significant improvements over benchmarks in localization accuracy, coverage robustness, and energy efficiency across both aerial and ground target-search scenarios.


\end{itemize}

The remainder is organized as follows: Sec. II reviews related work. Sec. III presents system model. Sec. IV details the optimization scheme. Sec. V presents simulations, and Sec.VI concludes the paper.

\section{Related Work}
This section reviews existing work in several interconnected research domains that collectively inspire our information-optimal UAV formation perception framework.

\subsection{Multimodal Cooperative Perception}
Multimodal cooperative sensing has emerged as a critical advancement to overcome the inherent limitations of single-platform perception systems in terms of coverage, accuracy, and environmental adaptability \cite{liuWhen2comMultiAgentPerception2020,xuV2XViTVehicletoEverythingCooperative2022}. Within the multi-UAV cooperative perception, Nitesh \textit{et al.} \cite{semenyukAdvancesUAVDetection2025} conducted an analysis of multimodal sensor integration techniques, which enhance detection precision, operational range, and overall system reliability. 
Tian \textit{et al.} \cite{tianUCDNetMultiUAVCollaborative2024} introduced a framework employing multi-view geometry and feature fusion to enable collaborative 3D object detection across multiple UAVs. Similarly, Qiao \textit{et al.} \cite{qiaoCoBEVFusionCooperativePerception2023} proposed a dual-window cross-attention mechanism that integrates LiDAR and camera features into a unified Bird's-Eye View representation. 
The development of standardized datasets has further accelerated research progress, with U2UData \cite{fengU2UDataLargescaleCooperative2024} establishing a benchmark featuring LiDAR point clouds, RGB/depth images, environmental sensing measurements, and 3D annotations for investigating multi-UAV multimodal perception and communication strategies.
Notably, the aforementioned approaches are data-driven, relying on deep neural networks trained on large annotated datasets. Their dependence on extensive training data and the associated inference overhead limits their applicability in data-scarce scenarios and on resource-constrained platforms.

\subsection{UAV Formation Design}

Constrained by sensor heterogeneity and geometric allocation, generating optimal UAV formation to enhance cooperative perception performance remains a pivotal challenge. 
In that, FIM analysis has been widely conducted for various perception techniques including time difference of arrival (TDOA), angle of arrival (AOA), received signal strength (RSS), and their hybrid frameworks. 
Panwar \textit{et al.} \cite{panwarOptimalSensorPlacement2022a} integrated time of arrival (TOA), RSS, and AOA measurements, deriving the Cramér--Rao Lower Bound (CRLB) expression for joint measurements to evaluate target localization accuracy. 
Hung \textit{et al.} \cite{hungOptimalSensingTracking2024} proposed a heterogeneous sensing multi-UAV tracking system modeling hybrid measurement configurations incorporating image sensors, TOA, and AOA to design optimal UAV formation.
Further, Sahu \textit{et al.} \cite{sahuOptimalSensorPlacement2022a} developed a unified optimization framework integrating Alternating Direction Method of Multipliers (ADMM) with Majorization-Minimization techniques, for optimal sensor placements. 
However, above FIM optimization typically yields idealized geometric allocations, which often results in severe FOV blind spots and inter-UAV communication interference.
Moreover, prior work cannot directly address the 3D placement of diverse sensors (cameras and LiDARs), failing to fuse the asymmetric information contributions along different spatial axes.

\subsection{Learning-Based Swarm Approaches}

Beyond the model-based methods reviewed above, recent research has explored learning-based paradigms for swarm coordination. Reinforcement learning (RL) has been increasingly adopted to derive adaptive multi-UAV formation and guidance policies in dynamic environments. For instance, Wu \textit{et al.} \cite{10492856} developed an attention-based MADDPG formation controller for hybrid UAVs, while Gavin \textit{et al.} \cite{10556867} applied multi-agent RL to drone guidance for cooperative triangulation. Such methods achieve strong adaptability through trial-and-error policy learning.
However, they generally require extensive offline training and incur non-trivial online inference costs, which are difficult to satisfy under the stringent size, weight, and power (SWaP) constraints of onboard UAV processors.
In contrast to these directions, our framework adopts a model-based, information-optimal formation that requires no deep learning training and relies solely on a preliminary target state estimate. It yields near-optimal formations with low computational latency, and jointly optimizes sensing accuracy, communication quality, and control efficiency within a unified geometric design.

\section{System Model}
\begin{figure*}[htbp] 
\centering
\includegraphics[width=0.95\textwidth]{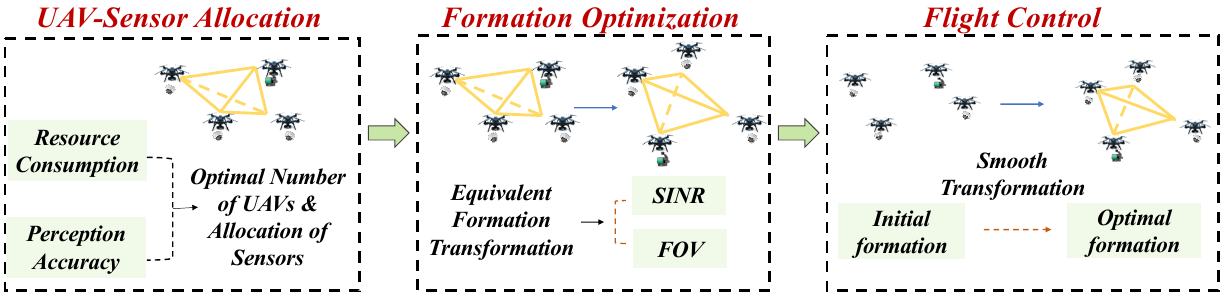} 
\caption{The illustration of the comprehensive framework for the cooperative perception of UAV swarms.}
\label{Flow}
\end{figure*}

This section presents our framework for UAV formation-based multimodal cooperative perception, structured around three sequential strategies. As shown in Fig.~\ref{Flow}, upon successful target identification, the UAV-sensor allocation and formation optimization strategies are executed only once: the former determines the number of UAVs and their sensor categories (LiDAR or Camera), and the latter enhances FOV coverage while preserving perception and communication quality. This one-shot initialization defines the mathematically optimal geometric allocation, avoiding redundant high-overhead optimization during cooperative flight.
In contrast, the flight control strategy operates continuously, iteratively updating the UAV kinematics in response to target maneuvers while maintaining the predefined optimal topology. 
This sequential approach holistically optimizes geometric allocation, sensor complementarity, communication, and flight energy. 

\subsection{System Assumption}
\begin{figure}
\centering
     \includegraphics[width=.31\textwidth]{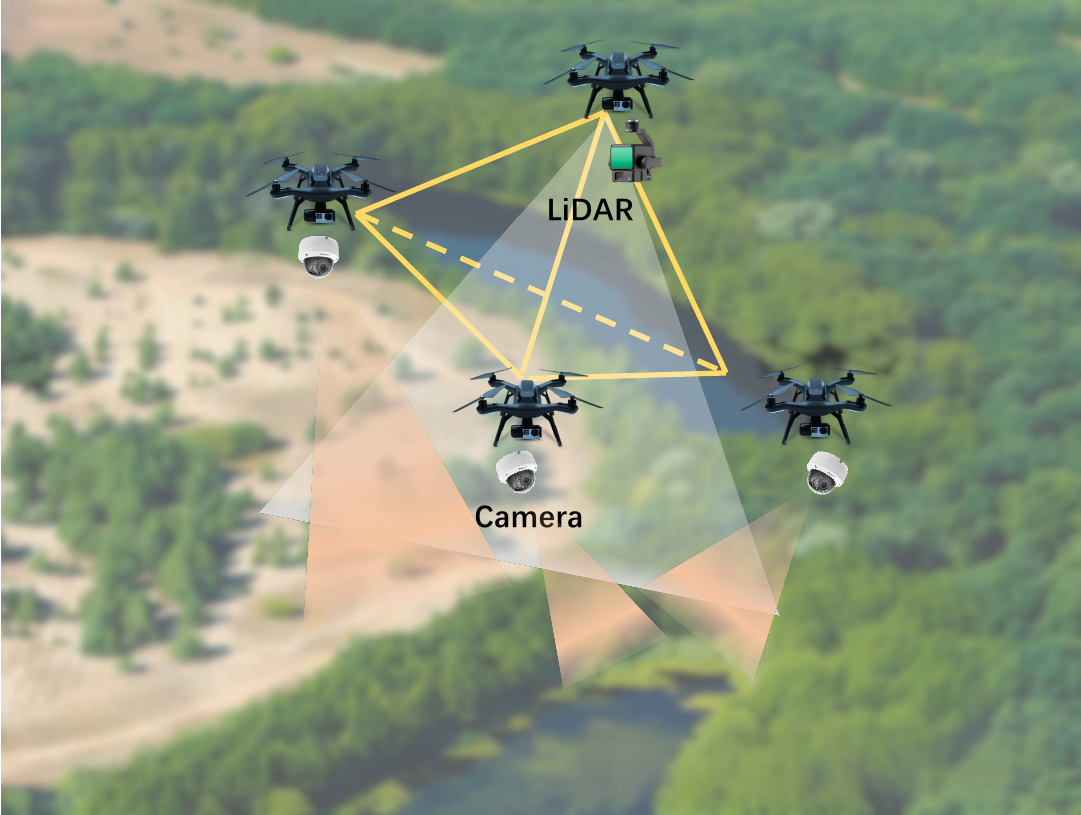} 
     \caption{UAV swarm-based SAR scenario.} 
\label{UAV_swarm_Scenario}
\end{figure}

The UAV formation for cooperative perception is demonstrated in Fig.~\ref{UAV_swarm_Scenario}, where heterogeneous UAVs collaboratively monitor an area for target detection and tracking.
These UAVs are equipped with complementary sensing modalities, some with image sensors (cameras) and others with point cloud sensors (LiDAR). 


The target, which may move on the ground or in the air, has a position vector ${\mathcal{P}_{tar}(t)} = [x_{tar}(t), y_{tar}(t), z_{tar}(t)]^T$ at time $t$. Each UAV $i$ has a position ${\mathcal{P}_{i}(t)} = [x_{i}(t), y_{i}(t), z_{i}(t)]^T$. We assume zero pitch angles during steady flight, and define the heading vector as $\mathbf{b}_i = [\cos\theta_i,\sin\theta_i,0]$, where $\theta_i$ denotes the yaw angle. Consequently, the relative 3D position vector is simply expressed as $\mathcal{P}_{i/tar}(t) = \mathcal{P}_{i}(t) - \mathcal{P}_{tar}(t)$, and its projection onto the horizontal X-Y plane is $\mathcal{P}_{i/tar}^{xy}(t) = [x_{i}(t)-x_{tar}(t), y_{i}(t)-y_{tar}(t), 0]^T$.

In practical SAR tasks, the target position $\mathcal{P}_{tar}(t)$ may not be directly observable. 
To address this, we formulate the cooperative perception as a predictor–corrector active sensing process. At planning step $t$, the UAV swarm employs a tracking filter (e.g., extended Kalman filter) to estimate the predicted target state $\hat{\mathcal{P}}_{tar}(t|t-1)$ based on previous observations. The formation geometry is then proactively optimized by relying on this prior state to acquire high-fidelity measurements, which subsequently refine the posterior state $\hat{\mathcal{P}}_{tar}(t|t)$.
Subsequently, we will elaborate on the UAV-sensor allocation, i.e., the formation allocation, for the multimodal cooperative perception.

\subsection{Multimodal Cooperative Perception}
This subsection quantifies multimodal cooperative perception via the FIM metric. We analyze two complementary modalities (vision cameras and point-cloud LiDAR) deriving their measurement models, Jacobians, and FIM formulations, and explicitly modeling each sensor's information contribution within the formation-aware framework.
The formulation supports optimal sensor allocation for localization accuracy while accounting for FOV limitations, measurement noise, and communication interference.

\subsubsection{Perception Measurement of Camera}
We construct the camera measurement model in three sequent steps: (i) transforming the target into the camera's local frame, (ii) applying the pinhole projection to obtain pixel coordinates, and (iii) computing the Jacobian for the FIM.

Here, the camera maps the 3D target position to the 2D image plane through a perspective projection. Upon this, we transform the global coordinates into the camera's local coordinate system. The pixel coordinates of the projection are denoted by ${g}_{\text{cam,i}} = [\mu_i, \nu_i]^T$, which is given as:
\begin{equation}
\begin{aligned}
{g}_{\text{cam,i}} = h_{\text{cam,i}}(\mathcal{P}_{tar},\mathcal{P}_i,\mathbf{b}_i) + \boldsymbol{\epsilon}_{\text{cam}},
\end{aligned}
\label{sadjfhaejr}
\end{equation}
\noindent where $\boldsymbol{\epsilon}_{\text{cam}}$ represents the measurement noise vector associated with the resolution level of the camera sensor and communication interference. 
$\mathbf{b}_i$ is the heading vector defined in Sec. III-A, which is determined by the UAV's yaw angle $\theta_i$.

$h_{\text{cam},i}(\cdot)$ is the projection function mapping the target position to the camera coordinate system. 
Assuming the camera's optical axis is aligned with the UAV's yaw angle $\theta_i$ and the pitch angle is zero. Hence, the target's coordinates in the camera's local frame $(X_c, Y_c, Z_c)^T$ are obtained via the rotation matrix $R_g^{c}$:
\begin{equation}
\begin{aligned}
\begin{bmatrix} X_c \\ Y_c \\ Z_c \end{bmatrix} 
&= R_g^{c} (\mathcal{P}_i-\mathcal{P}_{tar})\\
&= \begin{bmatrix}
-\sin\theta_i & \cos\theta_i & 0 \\
0 & 0 & -1 \\
\cos\theta_i & \sin\theta_i & 0
\end{bmatrix}
\begin{bmatrix}
x_i - x_{tar} \\
y_i - y_{tar} \\
z_i - z_{tar}
\end{bmatrix},
\end{aligned}
\end{equation}
\noindent where $Z_i = \cos\theta_i(x_i - x_{tar}) + \sin\theta_i(y_i - y_{tar})$ denotes the target depth along the camera optical axis.

Based on the standard pinhole camera model, the forward projection function $\pi(\cdot)$ transforms this 3D point to the pixel coordinates:
\begin{equation}
\begin{aligned}
&h_{\text{cam,i}}(\mathcal{P}_{tar},\mathcal{P}_i,\mathbf{b}_i) = 
\begin{bmatrix} \mu_i \\ \nu_i \end{bmatrix} = 
\begin{bmatrix} f_x \frac{X_c}{Z_i} + c_x \\[6pt] f_y \frac{Y_c}{Z_i} + c_y \end{bmatrix} \\
&= \begin{bmatrix}
-f_x \frac{\cos\theta_i(y_i - y_{tar}) - \sin\theta_i(x_i - x_{tar})}
        {\cos\theta_i(x_i - x_{tar}) + \sin\theta_i(y_i - y_{tar})} + c_x \\[10pt]
-f_y \frac{z_i - z_{tar}}
        {\cos\theta_i(x_i - x_{tar}) + \sin\theta_i(y_i - y_{tar})} + c_y
\end{bmatrix},
\end{aligned}
\end{equation}
\noindent where $f_x$ and $f_y$ are the focal lengths, and $c_x$ and $c_y$ are the principal point offsets.

Applying the chain rule to $h_{\text{cam,i}}$ with respect to the target state yields the $2 \times 3$ Jacobian:
\begin{equation}
\begin{aligned}
&\mathbf{O}_{\text{cam,i}} = \frac{\partial h_{\text{cam}}}{\partial \mathcal{P}_{tar}} \\
&=
\begin{bmatrix}
\dfrac{-f_x(y_i - y_{tar})}{Z_i^2} & \dfrac{f_x(x_i - x_{tar})}{Z_i^2} & 0 \\[12pt]
\dfrac{-f_y\cos\theta_i(z_i - z_{tar})}{Z_i^2} & \dfrac{-f_y\sin\theta_i(z_i - z_{tar})}{Z_i^2} & \dfrac{f_y}{Z_i}
\end{bmatrix}.
\end{aligned}
\label{csdhajckawyuv}
\end{equation}

Therefore, the FIM of the $i$-th UAV with a camera sensor is defined as:
\begin{equation}
\begin{aligned}
\mathcal{F}_{\text{cam,i}} = \mathbf{O}_{\text{cam,i}}^T \mathbf{Q}_{\text{cam}}^{-1} \mathbf{O}_{\text{cam,i}},
\end{aligned}
\label{cnaurbgaeur}
\end{equation}
\noindent where $\mathbf{Q}_{\text{cam},i}$ denotes the measurement covariance matrix of the camera sensor, which depends on the pixel resolution and the communication SINR.

\subsubsection{Perception Measurement of LiDAR}
\textcolor{black}{The LiDAR measurement comprises range, azimuth, and elevation, stacked into the full measurement vector and Jacobian. The range measurement is:}
\begin{equation}
\begin{aligned}
d_i = h_d(\mathcal{P}_{tar},\mathcal{P}_i) + \varepsilon_d,
\end{aligned}
\label{dajcvbueyrabdskv}
\end{equation}
\noindent where $\varepsilon_d$ is the measurement noise and $h_d(\mathcal{P}_{tar},\mathcal{P}_i)$ is the distance between the UAV $i$ and target:
\begin{equation}
\begin{aligned}
h_d(\mathcal{P}_{tar},\mathcal{P}_i) = \sqrt{(x_i - x_{tar})^2 + (y_i - y_{tar})^2 + (z_i - z_{tar})^2}.
\end{aligned}
\label{dsajcerkucbwasbc}
\end{equation}

Regarding 3D angle measurement, it is composed of the azimuth angle and the elevation angle of the target.
The azimuth angle measurement $\beta$ is expressed as:
\begin{equation}
\begin{aligned}
\beta_i = {h}_A(\mathcal{P}_{tar},\mathcal{P}_i) + \varepsilon_A ,
\end{aligned}
\label{sdjchuyawebciscn}
\end{equation}
\noindent where $\varepsilon_A$ is the measurement noise and $h_A(\mathcal{P}_{tar},\mathcal{P}_i)$ represents the azimuth angle difference, which is:
\begin{equation}
\begin{aligned}
h_A(\mathcal{P}_{tar},\mathcal{P}_i) = \tan^{-1}\left(\frac{y_i - y_{tar}}{x_i - x_{tar}}\right).
\end{aligned}
\label{ascnuewybvweba}
\end{equation}
In contrast, the elevation angle measurement $\delta$ is given as:
\begin{equation}
\begin{aligned}
\delta_i = {h}_p(\mathcal{P}_{tar},\mathcal{P}_i) + \varepsilon_p ,
\end{aligned}
\end{equation}
\noindent where $\varepsilon_p$ is the noise and $h_p(\mathcal{P}_{tar},\mathcal{P}_i)$ represents the elevation angle difference, which is:
\begin{equation}
\begin{aligned}
h_{p}(\mathcal{P}_{tar},\mathcal{P}_i) = \tan^{-1}\left(\frac{z_{i} - z_{tar}}{\sqrt{(x_{i} - x_{tar})^{2} + (y_{i} - y_{tar})^{2}}}\right).
\end{aligned}
\end{equation}

\noindent Based on the above analysis, the measurement of the LiDAR $h_\text{lidar}$ is composed by:
\begin{equation}
\begin{aligned}
h_{\text{lidar,i}}(\mathcal{P}_{tar},\mathcal{P}_i) 
&= 
\begin{bmatrix}
h_d(\mathcal{P}_{tar},\mathcal{P}_i) \\ 
h_A(\mathcal{P}_{tar},\mathcal{P}_i) \\
h_{p}(\mathcal{P}_{tar},\mathcal{P}_i)
\end{bmatrix} .
\end{aligned}
\end{equation}

Hereafter, the Jacobian matrix $\mathbf{O}_\text{lidar,i}$ of the LiDAR measurement for $i$-th UAV is:
\begin{equation}
\begin{aligned}
&\mathbf{O}_\text{lidar,i} = \frac{\partial h_{\text{lidar}}}{\partial \mathcal{P}_{tar}} \\
&=
\begin{bmatrix}
-\dfrac{x_i - x_{tar}}{d_i} & -\dfrac{y_i - y_{tar}}{d_i} & -\dfrac{z_i - z_{tar}}{d_i} \\[10pt]
\dfrac{\sin\beta_i}{d_i^{xy}} & -\dfrac{\cos\beta_i}{d_i^{xy}} & 0 \\[10pt]
\dfrac{(z_i - z_{tar})\cos\beta_i}{d_i^2} & \dfrac{(z_i - z_{tar})\sin\beta_i}{d_i^2} & -\dfrac{d_i^{xy}}{d_i^2}
\end{bmatrix},
\end{aligned}
\label{csdavuybearwscb}
\end{equation}

\noindent where $d_i^{xy}$ is the horizontal distance of the $i$-th UAV projected onto the tracking target, which is expressed as:
\begin{equation}
\begin{aligned}
d_i^{xy} = \sqrt{(x_i - x_{tar})^2 + (y_i - y_{tar})^2 }.
\end{aligned}
\label{kxcjvrudbcasioo}
\end{equation}

Finally, the FIM of LiDAR for $i$-th UAV is:
\begin{equation}
\begin{aligned}
\mathcal{F}_{\text{lidar,i}} = \mathbf{O}_{\text{lidar,i}}^T \mathbf{Q}_{\text{lidar}}^{-1} \mathbf{O}_{\text{lidar,i}},
\end{aligned}
\label{cnaurbgaeur}
\end{equation}

\noindent where $\mathbf{Q}_{\text{lidar}}$ is the measurement covariance matrix, which is related to the measurement and wireless transmission noise.
Subsequently, we elaborate on the resource constraints of the communication interference, flight energy, and sensor budgets.

As shown in Eq.~(\ref{csdhajckawyuv}) and Eq.~(\ref{csdavuybearwscb}), the Jacobian matrices $\mathbf{O}_{cam,i}$ and $\mathbf{O}_{lidar,i}$ inherently depend on the unknown true target position $\mathcal{P}_{tar}$. In our active sensing framework, these Jacobians are evaluated at the predicted state $\hat{\mathcal{P}}_{tar}(t|t-1)$ rather than the true state. 
This local approximation is a standard practice in information-driven sensor placement, enabling the swarm to maximize the expected information gain upon the current best estimate of the target. 
We note that this design choice explicitly scopes our framework to the active tracking regime. In the blind search phase, where no prior estimate exists, alternative exploration strategies (e.g., coverage-based or ergodic search) are required to establish initial target detection before our formation optimization can be applied.

For mutually independent measurements, the total FIM is equal to the sum of the individual FIMs. Assuming the formation includes $N_c$ camera-equipped UAVs and $N_l$ LiDAR-equipped UAVs, and the sensor noises are mutually independent, the global FIM with respect to the target position, $\mathcal{F}_{\text{total}}$, can be expressed as:

\begin{equation}
\begin{aligned}
    \mathcal{F}_{total} = \sum_{i=1}^{N_c} \mathcal{F}_{\text{cam, i}} + \sum_{j=1}^{N_l} \mathcal{F}_{\text{lidar,j}}.
\end{aligned}
\end{equation}

{\color{black}In practical deployment, our framework operates in the active tracking paradigm, assuming a preliminary target state is available via initial detection. Should this prediction deviate significantly, the swarm reverts to coverage-based search to re-acquire the target. To ensure mathematical tractability, we assume zero UAV pitch angles during steady-state tracking and an unobstructed line-of-sight (LoS). The LoS assumption holds well for aerial targets and open-terrain tracking, while terrain- and vegetation-induced occlusions are accommodated as model extensions in Sec.~IV-B.

The diagonal structure of the measurement covariance matrices ($\mathbf{Q}_{\text{cam}}$ and $\mathbf{Q}_{\text{lidar}}$) is justified by statistical independence at two levels: within each UAV, the measurement channels of a given sensor (the pixel coordinates ($\mu$, $\nu$) of the camera, and the range, azimuth, and elevation of the LiDAR) are driven by separate electronic processes. Across UAVs, the heterogeneous sensors are mounted on physically isolated airframes. 
The former underpins the FIM invariance under the equivalent transition in Sec.~IV-B, while the latter ensures the additivity of per-UAV FIMs. 
Multimodal data fusion additionally relies on tight time synchronization, precise camera--LiDAR extrinsic calibration, and reliable cross-view data association. 
Here, data association is guaranteed by the complementarity of visual semantics and geometric depth, and is further strengthened by the FOV-maximizing nature of our optimization, which can minimize occlusion-induced association failures.

These idealized baselines are reasonable in practice: the compact size of the UAV swarm renders air-to-air propagation delays negligible relative to the sensing intervals. 
Minor operational deviations, such as temporary pitch fluctuations or synchronization delays, can be modeled as zero-mean additive perturbations that inflate the diagonal entries of $\mathbf{Q}$. 
This degrades the achievable Cramér--Rao bound but preserves the FIM invariance property, as quantitatively validated in Sec.~V. More substantial violations, such as persistent strongly correlated noise, lie beyond the current scope.}

\subsection{Resource Consumption Analysis}
In the multimodal cooperative perception of the UAV swarm, the resource consumption mainly includes three aspects: communication consumption, flight maneuver energy consumption, and the inherent costs of heterogeneous sensors. Quantitative analysis of these costs is crucial for evaluating the feasibility of the system and optimizing the overall efficiency.

\subsubsection{SINR} In realistic dense UAV swarms, the air-to-air communication links suffer not only from distance-dependent attenuation but also from environmental shadowing and multipath fading. To reflect harsher communication conditions, the received signal power at UAV $j$ from UAV $i$ is modeled as:
\begin{equation}
p_{i,j}^{rec} = P_{tx} \rho_0 ||\mathcal{P}_{j}-\mathcal{P}_{i}||^{-\alpha} |h_{i,j}|^2,
\label{dsavhbauervbawusvbd}
\end{equation}

\noindent where $P_{tx}$ is the uniform transmission power, $\alpha$ is the path loss exponent, and $\rho_0$ is the reference path loss at 1 meter \cite{8894454}. Crucially, $|h_{i,j}|^2$ is the channel gain coefficient encompassing small-scale fading (e.g., Rayleigh or Rician fading) and log-normal shadowing, which introduces stochastic fluctuations into the link quality.

Consequently, the SINR of the communication link between the $i$-th and $j$-th UAV is expressed as:
\begin{equation}
\eta_{ij} = \frac{p_{i,j}^{rec}}{\sum_{k \in U, k \ne i,j} p_{k,j}^{rec} + \sigma^{2}},
\end{equation}
where $\sigma^2$ is the background noise power, $U$ is the set of UAVs engaged in cooperative sensing, and the denominator accounts for the severe co-channel interference generated by other closely operating UAVs in the swarm.

\subsubsection{Time-Frequency Communication Resource}
Regarding the $i$-th UAV, the time-frequency resource block for transmitting sensing data can be expressed as:
\begin{equation}
\begin{aligned}
{\mathscr{E}}_{i}^{comm} = {B}_i\cdot{T_i},
\end{aligned}
\end{equation}
\noindent where ${B}_i$ is the occupied bandwidth for $i$-th UAV's data transmission.
${T}_i$ represents the corresponding transmission duration for $i$-th UAV's, which can be regarded as the wireless channel occupation time.
Due to the inherently larger data volume of LiDAR-based point clouds compared to RGB images, LiDAR-equipped UAVs generally incur higher time-frequency resource block consumption with the prescribed transmission rate.



\subsubsection{Sensor Hardware Budgets}
Multimodal sensors incur distinct operational and manufacturing budgets. 
The hardware cost $C^{sensor}$ of onboard UAV sensors can generally be considered as a constant.
Additionally, the hardware cost $C_i^{lidar}$ of LiDAR is higher than that of cameras $C_i^{cam}$. Thus, $C_i^{lidar}>C_i^{cam}$.

\subsection{Properties of FIM}

To facilitate the analysis, feasible UAV deployment positions are modeled as a finite discrete set ${V}$, from which a subset $S$ is selected to form the active formation configuration. The objective function $f(S)$ evaluates this selected subset by maximizing the log-determinant of its FIM. To guarantee an optimal greedy solution, $f(S)$ satisfies two fundamental properties \cite{nakaiEffectObjectiveFunction2021a}: \textit{monotonicity} ($f(T) \geq f(S)$ for all $S \subseteq T \subseteq V$) and \textit{submodularity} ($f(S \cup \{v\}) - f(S) \geq f(T \cup \{v\}) - f(T)$ for any $v \in V \setminus T$). Monotonicity implies that adding a UAV never degrades perception, while submodularity encapsulates the diminishing marginal returns of adding new sensors. These properties guarantee that a greedy strategy yields a lower-bound performance of $(1-1/e) \approx 63.2\%$ relative to the global optimum.

\section{Formation Aware Strategies}
To address the cooperative perception problem, including sensing performance, communication, and flight energy constraint, we divide the UAV formation scheme into three subproblems: the UAV-sensor allocation, FOV-oriented formation optimization, and energy-efficient flight control problems.

\subsection{UAV-Sensor Allocation}
The first subproblem aims to determine the optimal number of UAVs and the multimodal sensor allocation (camera and LiDAR) within the swarm. The objective is to maximize the collaborative sensing accuracy for target tracking, while considering the diminishing reward of new node incorporation and the associated costs of communication overhead and hardware deployment.

We adopt the D-optimality (Determinant-optimality) criterion from optimal experimental design, which leads to the following objective function \cite{6360016Morbidi}:
\begin{equation}
f(U) = \log \det \mathcal{F}(U),
\end{equation}
\noindent where $U \subseteq V$ is the subset of the candidate set ${V}$. $\mathcal{F}(U)$ is the FIM for the selected allocation. Crucially, the log-determinant of the FIM has been proven to satisfy both monotonicity and submodularity \cite{6360016Morbidi, yamadaFastGreedyOptimization2021}, thereby validating the theoretical applicability of the proposed greedy selection strategy.
To balance sensor performance of different UAVs with practical resource constraints, we augment the objective with penalty terms for communication time-frequency resource and hardware costs:

\begin{equation}
\mathcal{J}(U) = \log \det \mathcal{F}(U) - w_1 \sum_{v \in U} {\mathscr{E}}_{v}^{comm} - w_2 \sum_{v \in U} C_v^{\text{sensor}},
\label{asxluibweugxubwk}
\end{equation}
\noindent where $\mathscr{E}_v^{\text{comm}}$ denotes the time-frequency resources block of UAV $v$, $C_v^{\text{sensor}}$ is the sensor hardware costs, and $w_1$, $w_2$ are weight coefficients that regulate the trade-off between perception performance and resource utilization.
Based on the above analysis, we provide the optimization model for formation cooperative perception as:

\begin{equation}
\begin{aligned}
\text{(P1):} \quad \max_{U \subseteq V} \quad  
&   \mathcal{J}(U) \\
\text{s.t.} \quad &  C_1:|U| \leq N_{\text{max}} \\
&  C_2:d_{\text{min}} \leq d_v \leq d_{\text{max}}, \quad \forall v \in U \\
&  C_3:\beta_{\text{min}} \leq \beta_v \leq \beta_{\text{max}}, \quad \forall v \in U \\
&  C_4:\delta_{\text{min}} \leq \delta_v \leq \delta_{\text{max}}, \quad \forall v \in U \\
&  C_5:s_v \in \{\text{Camera}, \text{LiDAR}\}, \quad \forall v \in U
\end{aligned},
\label{whole_sys_model}
\end{equation}

\noindent in which $C_1$ is the maximum number of deployed UAVs according to practical conditions and coordination capabilities. $C_2$ gives the distance range constraints, designed to ensure a safe separation from the target while simultaneously keeping the UAV within the sensor's effective perception coverage. 
$C_3$ and $C_4$ specify the azimuth and elevation angle constraints, respectively, guaranteeing that the target remains within the sensor's effective FOV. 
And $C_5$ represents the available sensor category, i.e., the camera and LiDAR.

Notably, the monotonicity and submodularity properties of the FIM enable us to employ efficient greedy selection algorithms with theoretical optimal value guarantees. 
The marginal gain of incorporating an additional UAV $v$ into an existing set $U_j$ that contains $j$ number of UAVs is given as:

\begin{equation}
\begin{aligned}
\Delta(v | U_{j}) = f(U_{j} \cup v) - f(U_{j}).
\end{aligned}
\label{ksdjvbguearbvsx}
\end{equation}

\textcolor{black}{Moreover, to implement the greedy selection algorithm, we introduce a utility function that incorporates both marginal gains and system costs:}
\begin{equation}
\begin{aligned}
G(v) = \Delta(v | U_{j}) - w_1 \mathscr{E}_v^{comm} - w_2 C_v^{sensor}.
\end{aligned}
\label{casiuvbaeruybcnx}
\end{equation}

The core of our greedy selection strategy involves iteratively choosing UAV spatial allocations from a finite candidate set ${V}$, constructed through discretization of the solution space encompassing all feasible UAV positions and sensor combinations. 
Specifically, the candidate set ${V}$ is given as:

Spatial Discretization: We model the airspace around the target in a target-centered spherical coordinate system, discretizing the radial distance $d$ at intervals $\Delta d$ within $[d_{\text{min}}, d_{\text{max}}]$ (bounded by the effective sensing range) and uniformly sampling the azimuth $\beta$ and elevation $\delta$ at resolutions $\Delta\beta$ and $\Delta\delta$. It yields a grid of candidate points $p_{\text{candidate}} = (d, \beta, \delta)$.

Sensor Assignment: For each candidate position, we generate two configurations---one camera and one LiDAR---which doubles the candidate set but enables optimal modality selection under information and hardware constraints.

View Direction: The yaw $\theta_i$ is set by aligning the sensor's optical axis toward the target, making it a deterministic function of the relative position rather than an independent variable, which reduces the search dimensionality.

Algorithm~\ref{Alg1} summarizes the complete greedy selection procedure. The algorithm terminates when the maximum utility gain $G(v_t)$ falls below a predefined threshold $\rho$, indicating that the inclusion of additional UAVs no longer offsets their associated communication and hardware costs. This termination criterion, together with the submodular property of the objective function, guarantees a $(1-1/e)$-approximation to the optimal solution. 

\textcolor{black}{More importantly, the computational complexity of this procedure is rigorously bounded by $\mathcal{O}(M|V|)$, where $M$ denotes the final number of selected UAVs and $|V|$ is the size of the discrete candidate set. In each of the $M$ iterations, the algorithm evaluates the marginal gain for the remaining candidate allocations. Given that the practical swarm size $M$ is typically small, the algorithm effectively scales linearly with the search space size. This manageable linear complexity makes the proposed greedy strategy well suited for real-time onboard execution on resource-constrained UAV platforms in time-critical SAR tasks.}



\begin{algorithm}
\caption{UAV-sensor Greedy Selection} \label{Alg1}
	
\textbf{Input}: Candidate set ${V}$, cost coefficients $w_1, w_2$, time-frequency resource $\mathscr{E}$, hardware budgets $C^{sensor}$

\textbf{Output}: UAV-sensor allocation $U_M$

$U_0 \gets \emptyset$; $M \gets 0$

\While{true}
{
    $M \gets M + 1$
    
    \For{each $v \in {V} \setminus U_{M-1}$}
    {
        $\Delta(v | U_{M-1}) \gets \log\det \mathcal{F}(U_{M-1} \cup v) - \log\det \mathcal{F}(U_{M-1})$
        
        $\text{G}(v) \gets \Delta(v | U_{M-1}) - w_1 \mathscr{E}_v^{comm} - w_2 C_v^{sensor}$
    }
    
    $v_t \gets \arg\max_{v \in {V} \setminus U_{M-1}} G(v)$
    
    \If{$G(v_t) \leq$ termination threshold $\rho$}
    {
        \textbf{break}
    }
    
    $U_M \gets U_{M-1} \cup \{v_t\}$
}

\end{algorithm}

\subsection{Formation Optimization}



\textcolor{black}{Upon the UAV-sensor allocation through Alg.~\ref{Alg1}, we now address the optimal spatial formation to maximize the FOV coverage without FIM loss. 
While the selected UAVs and sensors provide theoretically optimal target localization accuracy, their initial geometric allocation may exhibit significant FOV gaps and communication interference that compromise operational robustness in dynamic SAR scenarios. 
This subsection introduces an equivalent transition framework that preserves the fundamental perception accuracy (the FIM determinant) while strategically reconfiguring UAV positions to maximize FOV coverage completeness and communication quality. }

\begin{figure}[htbp]
\centering
     \includegraphics[width=.3\textwidth]{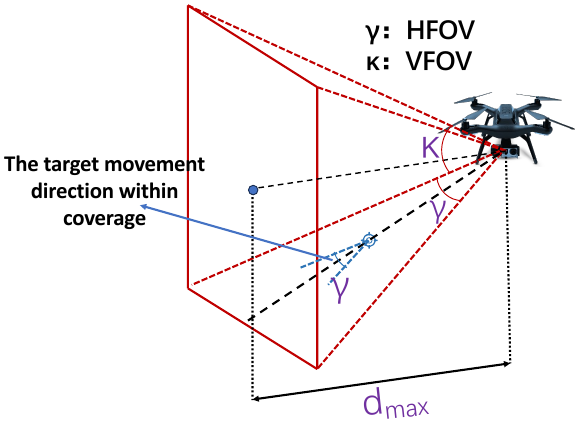} 
     \caption{Demonstration of FOV.} 
\label{FOV}
\end{figure}

To ensure continuous target tracking during surveillance operations, FOV coverage constitutes a critical performance metric beyond mere perception accuracy. An optimal formation must not only maintain target observability but also provide spatially balanced observation perspectives to mitigate occlusion risks and accommodate potential target maneuvers. We model each UAV's FOV coverage as a rectangular pyramidal frustum determined by three geometric parameters: maximum perception range $d_{\text{max}}$, horizontal FOV (HFOV) $\gamma$, and vertical FOV (VFOV) $\kappa$. \textcolor{black}{Crucially, $d_{\text{max}}$ denotes the effective sensing range. It is physically constrained by the minimum resolution that the onboard sensors require relative to the target's physical dimensions. Thus, the selection of $d_{\text{max}}$ implicitly accounts for the expected sizes of characteristic objects, such as humans or vehicles, within the specific SAR task context. The current FOV model assumes an unobstructed LoS; terrain- and vegetation-induced occlusions can be incorporated by introducing a probabilistic LoS coefficient into the global FIM and by inflating the covariance $\mathbf{Q}$, which we leave as a model extension. Fig.~\ref{FOV} presents the pyramidal frustum-based FOV demonstration.}

{\color{black}To capture fine-grained perception gains, we measure the coverage quality using a \textit{quality-weighted coverage metric} that explicitly reflects viewing geometry and observation range. Then, we discretize the horizontal space surrounding the target into $\mathcal{N}$ equally spaced azimuth directions represented by unit vectors $\mathbf{n}_k = [\cos\phi_k, \sin\phi_k, 0]^T$, where $\phi_k = 2\pi k/\mathcal{N}$ for $k=0, 1, \cdots, \mathcal{N}-1$. The number of sampled directions $\mathcal{N}$ is determined by the required angular resolution.
Denote $\varphi_{i,k}$ the angular deviation between the discrete direction $\mathbf{n}_k$ and the $i$-th UAV's horizontal observation vector $\mathcal{P}_{i/tar}^{xy}$:
\begin{equation}
\varphi_{i,k} = \arccos \left( \frac{\mathbf{n}_k \cdot \mathcal{P}_{i/tar}^{xy}}{\|\mathcal{P}_{i/tar}^{xy}\|} \right).
\label{angular_deviation}
\end{equation}

In practical optical and LiDAR sensors, perception quality degrades gracefully as the target deviates from the optical center and as the relative distance increases. Thus, we define a continuous quality-weighted coverage intensity $q(i, k)$ contributed by UAV $i$ to direction $\mathbf{n}_k$:
\begin{equation}
q(i, k) = 
\begin{cases} 
\exp\left(-\frac{\varphi_{i,k}^2}{2\sigma_{\gamma}^2}\right) \cdot \frac{d_{\text{ref}}^2}{\|\mathcal{P}_{i/tar}\|^2}, & 
\begin{aligned}
&\text{if } \varphi_{i,k} \le \frac{\gamma_i}{2} \\
&\text{and } \|\mathcal{P}_{i/tar}\| \le d_{\text{max}}
\end{aligned} \\ 
0, & \text{otherwise} 
\end{cases},
\label{quality_intensity}
\end{equation}
where $\sigma_{\gamma}$ controls the Gaussian sensitivity of perception degradation off the optical axis, and $d_{\text{ref}}$ is a normalizing reference distance. Physically, this intensity metric models the sensor-specific detection reliability and spatial resolution capacity, explicitly capturing the confidence drop at the periphery of the FOV. For each discrete direction $\mathbf{n}_k$, the total coverage intensity is the superposition of contributions from all $M$ deployed UAVs (obtained from Alg.~\ref{Alg1}):
\begin{equation}
\Phi(\mathbf{n}_k) = \sum_{i=1}^M q(i, k).
\label{total_intensity}
\end{equation}

To explicitly penalize completely uncovered directions (blind spots) and promote coverage integrity, we introduce a penalty multiplier $\xi$:
\begin{equation}
\xi = 1-\frac{\mathcal{N}_{\text{uncoveblack}}}{\mathcal{N}},
\label{penalty_term}
\end{equation}
where $\mathcal{N}_{\text{uncoveblack}} = \sum_{k=0}^{\mathcal{N}-1} \mathbb{I}(\Phi(\mathbf{n}_k) = 0)$ denotes the count of directions receiving absolutely zero coverage, with $\mathbb{I}(\cdot)$ being the indicator function. Finally, the comprehensive FOV coverage metric $\Gamma$ is formulated as:
\begin{equation}
\Gamma = \xi \cdot \left( \frac{1}{\mathcal{N}} \sum_{k=0}^{\mathcal{N}-1} \Phi(\mathbf{n}_k) \right).
\label{comprehensive_gamma}
\end{equation}

Here the $1/\mathcal{N}$ normalization yields an average quality-weighted intensity, while the penalty multiplier $\xi \in [0, 1]$ scales down the score whenever blind spots exist, driving the optimization toward seamless spatial coverage.}

Given the initial UAV set $U_M$ from Alg.~\ref{Alg1}, we formulate the formation optimization problem to maximize FOV coverage while preserving perception performance.
{\color{black}Crucially, we leverage the geometric symmetry of the observation manifold to perform formation transition. A central symmetry transition relocates a UAV to the antipodal position relative to the target, achieved by adding $\pi$ to the azimuth angle and negating the elevation angle. Such a transition alters the signs of specific rows in the observation Jacobian $\mathbf{O}_i$, represented by a diagonal sign matrix $\mathbf{S}_i$ containing $\pm 1$ on its diagonal. The transformed Jacobian is $\mathbf{O}_i' = \mathbf{S}_i \mathbf{O}_i$, yielding the updated FIM as:
\begin{equation}
\mathcal{F}_i' = (\mathbf{S}_i \mathbf{O}_i)^T \mathbf{Q}^{-1} (\mathbf{S}_i \mathbf{O}_i) = \mathbf{O}_i^T (\mathbf{S}_i^T \mathbf{Q}^{-1} \mathbf{S}_i) \mathbf{O}_i.
\end{equation}

The invariance of the aggregated FIM relies on the two independence assumptions established in Sec.~III-B. First, the measurement covariance $\mathbf{Q}$ is strictly diagonal; since diagonal matrices commute, $\mathbf{S}_i^T \mathbf{Q}^{-1} \mathbf{S}_i = \mathbf{Q}^{-1}$ holds ($\mathbf{S}_i^T \mathbf{S}_i = \mathbf{I}$), preserving the per-UAV FIM ($\mathcal{F}_i' = \mathcal{F}_i$). Second, statistically independent cross-UAV noise ensures the aggregated FIM remains the linear sum of per-UAV FIMs, preserving global invariance~\cite{zhaoOptimalSensorPlacement2013}. Consequently, this central-symmetry reconfiguration is a mathematically exact operation that improves FOV coverage and communication topology without sacrificing the CRLB. If these conditions are violated (non-diagonal covariance, correlated noise, or large calibration errors), the exact cancellation degrades into a near-optimal heuristic. Hence, the optimization problem is:}



\begin{equation}
\begin{aligned}
(\text{P2}): \quad & \max_{\mathcal{P}_i \in \Omega_{sym}(U_M)} \Gamma \\
\text{s.t.} \quad & \eta_i^{SINR} \ge \eta_{min}^{SINR}, \quad \forall i \in U_M
\end{aligned},
\label{asdvbewkusbvwuiebd}
\end{equation}

\noindent where $\eta^\text{SINR}_{\text{min}}$ is the minimum required SINR ratio for reliable wireless communication. $\Omega_{sym}(U_M)$ is the discrete space of equivalent formations generated by applying the central symmetry transition to any subset of UAVs in $U_M$.

To address the combinatorial complexity of evaluating all $2^M$ possible flip allocations for $M$ UAVs, we develop an efficient search strategy based on azimuth sector partitioning. 
The algorithm first computes the yaw angle $\theta_i$ of each UAV with respect to the target. Since each UAV is assumed to face the target, the yaw angle is determined by the bearing from 
the UAV to the target:
\begin{equation}
    \theta_i = \operatorname{atan2}\!\left( y_{\mathrm{tar}} - y_i,\; x_{\mathrm{tar}} - x_i \right),
    \label{zfjvzksvjxkvnf}
\end{equation}

\noindent where $\operatorname{atan2}(y, x)$ denotes the two-argument arctangent function, which returns the angle of the vector $(x,\, y)$ in the full range $(-\pi,\, \pi]$ and correctly handles all quadrants. These angles are then partitioned into $K$ sectors according to:

\begin{equation}
\begin{aligned}
\text{sector}_i = \lfloor \frac{\theta_i}{2\pi/K} \rfloor.
\end{aligned}
\label{sdakfvhbuefbve}
\end{equation}

\noindent Within sectors containing multiple UAVs (typically exhibiting high coverage overlap), we prioritize flipping operations for UAVs that yield the greatest marginal improvement in $\Gamma$ while satisfying the SINR constraint. 

Alg.~\ref{Alg2} details this formation optimization procedure, which systematically evaluates candidate flips while pruning the search space through sector-based constraints. This approach efficiently identifies the UAV formation that maximize FOV coverage completeness without compromising perception accuracy or communication reliability.


\begin{algorithm}
\caption{FOV-oriented Formation Optimization} \label{Alg2}
	
\textbf{Input}: Formation ${U}$ from Alg.~\ref{Alg1}, target position $\mathbf{p}_t$


\textbf{Output}: Optimal formation $U^*$

$U^* \gets U$, $\mathscr{T}^* \gets \Gamma(U)$

\For{each $u_i \in U$}
{
    $\theta_i \gets  \operatorname{atan2}\!\left( y_{\mathrm{tar}} - y_i,\; x_{\mathrm{tar}} - x_i \right)$,
    
    $\text{sector}_i \gets \lfloor \frac{\theta_i}{2\pi/K} \rfloor$ 
}

\For{each $\text{sector}_k$ with multiple UAVs}
{
    \For{each $u_i \in sector_k$}
    {
        $U' \gets \text{Flip}(U, u_i)$, 
        
        $\mathscr{T}' \gets \Gamma(U')$
        
        \If{$\eta^\text{SINR}(U') \geq \eta^\text{SINR}_{\text{min}}$ and $\mathscr{T}' > \mathscr{T}^*$}
        {
            $U^* \gets U'$
                
            $\mathscr{T}^* \gets \mathscr{T}'$
        }
    }
}

\end{algorithm}

\subsection{Flight Control}



Building upon the optimal FOV formation derived in Sec. IV-B, this subsection addresses the challenge of transitioning the UAV swarm from its initial position to the desired geometric formation $\mathbf{\mathcal{P}}^* = [\mathcal{P}_1^*, \mathcal{P}_2^*, \cdots, \mathcal{P}_m^*]^T$ while ensuring energy efficiency, trajectory smoothness, and asymptotic stability. 
\textcolor{black}{Conventional control approaches often prioritize rapid convergence at the expense of energy consumption and maneuver aggressiveness. In contrast, we develop a Lyapunov-based distributed control framework that systematically balances formation accuracy with operational efficiency.} The proposed control architecture incorporates a logarithmic potential field to bound control authority during large formation errors, thereby preventing excessive maneuvers while guaranteeing global asymptotic stability. 



At the beginning, we model each UAV's dynamics using a double integrator system:
\begin{equation}
\begin{aligned}
\dot{\mathcal{P}}_i(t) &= \mathbf{v}_i(t) ,
\end{aligned}
\end{equation}
\begin{equation}
\begin{aligned}
m_i \dot{\mathbf{v}_i}(t) &= \mathbf{u}_i(t),
\end{aligned}
\end{equation}
where $\mathcal{P}_i$, $\mathbf{v}_i$, and $m_i$ represent the position, velocity, and mass of the $i$-th UAV, respectively, and $\mathbf{u}_i(t)$ denotes the control input.
Additionally, the desired formation is characterized by target relative displacements $d_{ij}$ between neighboring UAV pairs $(i,j)$, satisfying $\mathcal{P}_i^* - \mathcal{P}_j^* = d_{ij}$ where $d_{ij} = -d_{ji}$. Then, the formation error between UAVs $i$ and $j$ is defined as:
\begin{equation}
\begin{aligned}
\mathbf{e}_{ij}(t) = \mathcal{P}_i(t) - \mathcal{P}_j(t) - d_{ij},
\end{aligned}
\label{asclduvybkawsbvcs}
\end{equation}

\noindent where $\mathcal{P}_i^*$ indicates the desired position of the $i$-th UAV. The formation objective requires $\mathbf{e}_{ij} \to 0$
for all connected pairs. The communication topology is denoted by a connected undirected graph $\mathcal{G}$, with $\mathcal{M}_i$, the neighbor set of UAV $i$.

Upon these dynamics, we design the control scheme in a leader-follower structure and then verify its stability via a Lyapunov argument. Specifically, a designated leader UAV $L$ incorporates trajectory tracking capability, while the follower UAVs execute distributed formation control.
The proposed control scheme is formulated as:
\begin{equation}
\mathbf{u}_i =
\begin{cases}
\begin{aligned}
& -k_1 \sum_{j \in \mathcal{M}_L} \frac{\mathbf{e}_{Lj}}{1 + \|\mathbf{e}_{Lj}\|^2} \\[-2pt]
& - k_2 \mathbf{v}_L - k_p (\mathcal{P}_L - \mathcal{P}_L^{\text{des}}),
\end{aligned} & i=L, \\[20pt]
-k_1 \sum_{j \in \mathcal{M}_i} \frac{\mathbf{e}_{ij}}{1 + \|\mathbf{e}_{ij}\|^2} - k_2 \mathbf{v}_i, & i \neq L,
\end{cases}
\label{sadvaunrbvwuinx}
\end{equation}
\noindent where $\mathcal{P}_L^{\text{des}}$ represents the leader's desired absolute position (determined by the tracked target), and $k_1, k_2, k_p > 0$ are positive control gains.

The stability of the closed-loop system is established using Lyapunov theory. Consider the candidate Lyapunov function: 

\begin{equation}
\begin{aligned}
\mathcal{L} &= \sum_{(i,j) \in \mathcal{G}} \ln (1 + \|\mathbf{e}_{ij}\|^2) + \sum_i \frac{1}{2} \|\mathbf{v}_i\|^2  \\&+k_3 \|\mathcal{P}_L -\mathcal{P}_L^{\text{des}}\|^2.
\end{aligned}
\label{eiubgvsajcw}
\end{equation}

This positive definite function combines three key components: (i) a logarithmic potential field $\ln(1 + \|\mathbf{e}_{ij}\|^2)$ that bounds control authority during large formation errors, preventing excessive maneuvers; (ii) a velocity damping term $\|\mathbf{v}_i\|^2$ that suppresses oscillations and ensures smooth velocity profiles; and (iii) a trajectory tracking component $k_3 \|\mathcal{P}_L -\mathcal{P}_L^{\text{des}}\|^2$ for the leader UAV $L$. Together, these elements synergistically balance convergence speed with energy efficiency while guaranteeing formation accuracy. Hereafter, differentiating Eq.~(\ref{eiubgvsajcw}) with respect to time yields:
\begin{equation}
\begin{aligned}
\dot{\mathcal{L}} &= \sum_{i=1}^{M} \left( \sum_{j \in \mathcal{M}_i} \frac{2\mathbf{e}_{ij}}{1 + \|\mathbf{e}_{ij}\|^2} \right)^T \mathbf{v}_i \\&+ \sum_{i=1}^{M} \mathbf{v}_i^T \mathbf{u}_i 
+k_3 (\mathcal{P}_L - \mathcal{P}_L^{\text{des}})^T \mathbf{v}_L.
\end{aligned}
\label{iusdgvuserubdv}
\end{equation}

 Substituting the control scheme (Eq.~(\ref{sadvaunrbvwuinx}) and (\ref{sadjvaukweyfbdwauiq})) into the system dynamics (Eq.~(\ref{iusdgvuserubdv})) with $k_1 = 4$, $k_p = k_3 $, we obtain:
\begin{equation}
\begin{aligned}
\dot{\mathcal{L}} = -k_2 \sum_{i=1}^{M} \|\mathbf{v}_i\|^2 \leq 0.
\end{aligned}
\label{sdaviudbveurs}
\end{equation}

By LaSalle's invariance principle \cite{lasalle1960some}, the system is globally asymptotically convergent, ensuring all UAVs converge to their designated positions in the desired formation. 
The logarithmic potential field in the Lyapunov function specifically optimizes trajectory smoothness by preventing control saturation during large formation errors, thereby minimizing energy-intensive maneuvers while maintaining precise formation tracking. This design achieves an optimal balance between formation accuracy, trajectory smoothness, and energy efficiency for UAV swarm operations.


\section{Numerical Simulation}

This section numerically validates the proposed framework along three modules: the UAV-sensor allocation, the FOV-oriented optimal formation, and energy-efficient flight control. 
It quantifies its improvements over conventional benchmarks under practical constraints of communication interference, energy limitations, and hardware budgets.

{\color{black}To ensure reproducibility, the critical simulation settings are listed in Tab.~\ref{Simulation_parameters}. The key parameters governing the optimization, communication, and control layers are explicitly defined. The objective weighting coefficients ($w_1, w_2$) prioritize perception information gains during initial iterations while penalizing redundant hardware allocation. For the greedy selection algorithm, the termination threshold ($\rho$) is calibrated to balance perception accuracy with system overhead. Furthermore, the flight control gains ($k_1, k_2, k_p$) were optimized via grid search to guarantee smooth formation transitions.}


\begin{table}[htbp]
\centering
\caption{Key Simulation Parameters}
\label{Simulation_parameters}
\begin{tabular}{llc}
\toprule
\textbf{Parameter Description} & \textbf{Symbol} & \textbf{Value} \\
\midrule
{\color{black}Maximum candidate UAVs} & {\color{black}$M_{\max}$} & {\color{black}10} \\
{\color{black}Target velocity vector} & {\color{black}$\mathbf{v}_{\text{tar}}$} & {\color{black}$[0.5, 0.3, 0.0]^T$ m/s} \\
{\color{black}FOV azimuth evaluation step} & {\color{black}$\Delta\phi$} & {\color{black}$10^\circ$ (36 directions)} \\
{\color{black}Carrier frequency} & {\color{black}$f_c$} & {\color{black}2.4 GHz} \\
{\color{black}Transmit power} & {\color{black}$P_t$} & {\color{black}20 dBm} \\
{\color{black}Minimum SINR threshold} & {\color{black}$\eta_{min}^{SINR}$} & {\color{black}10 dB} \\
Noise power & $\sigma^2$ & -110 dBm \\
LiDAR measurement covariance & $\mathbf{Q}_{\text{lidar}}$ & $\text{diag}(0.1, 0.02, 0.015)$ \\
Camera measurement covariance & $\mathbf{Q}_{\text{cam}}$ & $\text{diag}(6, 6)$ \\
Horizontal FOV limit & $\gamma$ & $50^\circ$ \\
Vertical FOV limit & $\kappa$ & $40^\circ$ \\
\midrule
Objective weighting coefficients & $w_1, w_2$ & 0.18, 0.20 \\
Greedy termination threshold & $\rho$ & 0.17 \\
        \textcolor{black}{Coverage metrics} & \textcolor{black}{$\sigma_\gamma, d_{\text{ref}}$} & \textcolor{black}{$25^\circ$, 10 m} \\
Flight control gains & $k_1, k_2, k_p$ & 4.0, 1.5, 10.0 \\

\bottomrule
\end{tabular}
\end{table}


\subsection{UAV-Sensor Allocation}

We implemented the proposed greedy algorithm detailed in Sec. IV-A to determine the optimal allocation of UAVs and sensor modalities. The objective function maximizes $\log\det \mathcal{F}(C)$, while incorporating penalty terms for resource consumption $\mathscr{E}^{comm}$ and hardware budgets $C^{sensor}$. 

\begin{figure}
\centering
     \includegraphics[width=.35\textwidth]{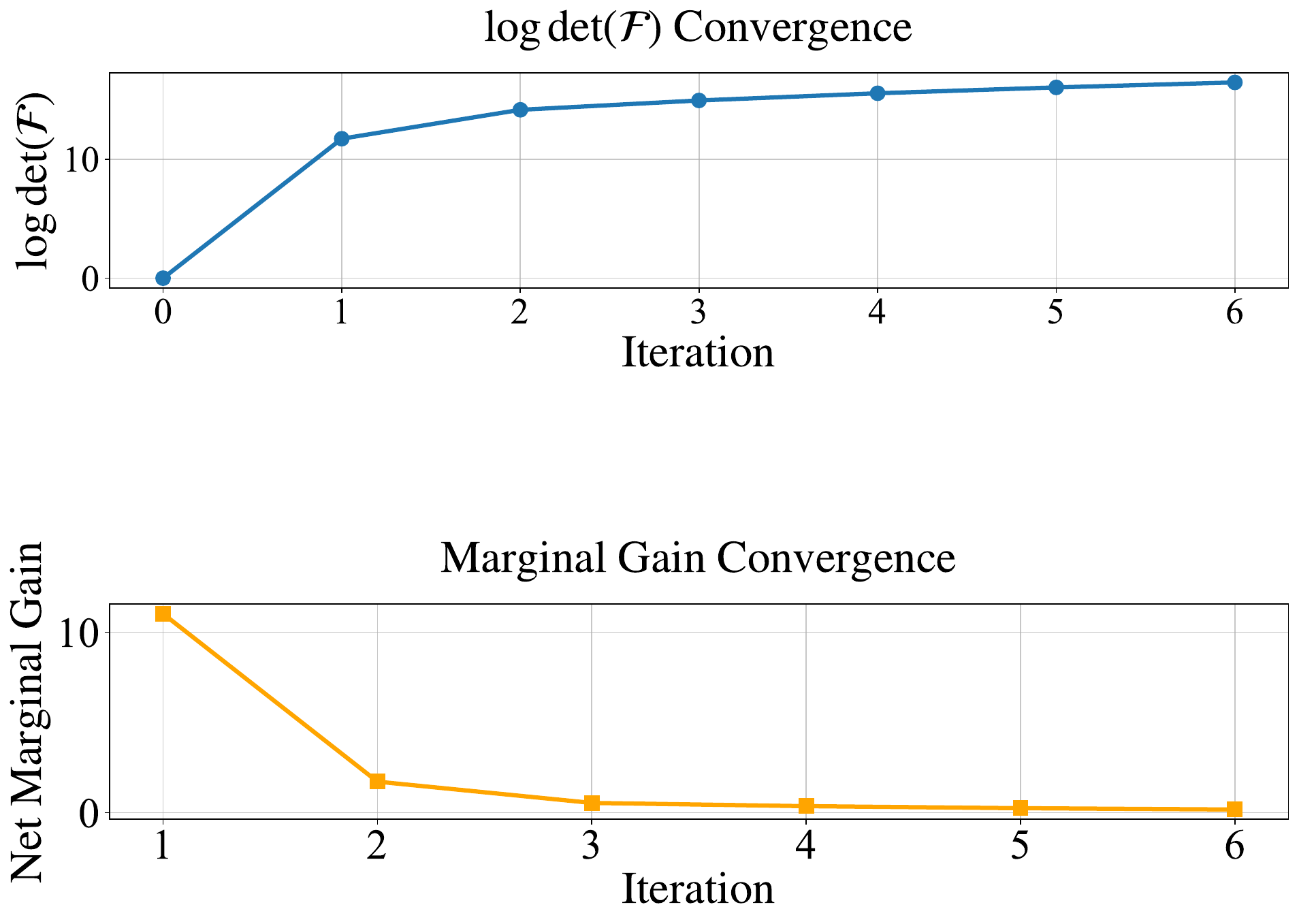} 
     \caption{Convergence of the greedy UAV-sensor allocation. Top: $\log\det(\mathcal{F})$ saturates as the number of UAVs increases. Bottom: the net marginal gain diminishes accordingly, confirming the submodularity of the objective.} 
\label{Q1_curves}
\end{figure}

Fig.~\ref{Q1_curves} shows the convergence of the Alg.~\ref{Alg1}. As expected from the submodular property of the FIM determinant, the perception accuracy shows diminishing returns as the number of UAVs of the formation increases. This result justifies the termination criterion for the greedy algorithm (Alg.~\ref{Alg1}). 
Mentioned that Fig.~\ref{Alg1} yields an optimal UAV-sensor allocation comprising by four camera-equipped UAVs and two LiDAR-equipped UAUs. 
The corresponding formation geometry is visualized in Fig.~\ref{Q1_formation}, with parameters provided in Tab.~\ref{table_uav_config1}.

\begin{figure}
\centering
     \includegraphics[width=.35\textwidth]{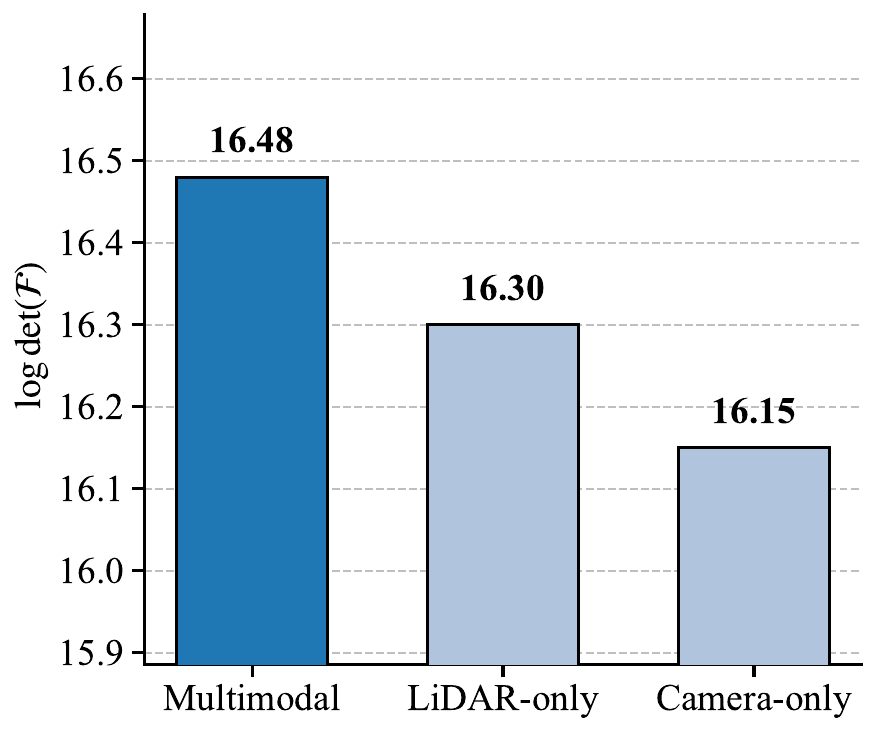} 
     \caption{Performance between multi- and single-modal formation.} 
\label{Q1_bars}
\end{figure}

\textcolor{black}{Fig.~\ref{Q1_bars} provides the performance comparison between multimodal and single-modal formation by applying Alg.~\ref{Alg1}. Here, the multimodal case allows both cameras and LiDARs, whereas the single-modal case restricts UAVs to those equipped only with cameras or LiDARs.
The optimal multimodal formation attains a $\log\det(\mathcal{F})$ of $16.48$ with only 6 UAVs (2 LiDARs, 4 cameras). However, the LiDAR-only and camera-only allocations reach $16.30$ (4 LiDARs) and $16.15$ (7 cameras). This shows that multimodal formation can improve perceptual accuracy through the sensor diversity.}

\begin{figure}
\centering
     \includegraphics[width=.35\textwidth]{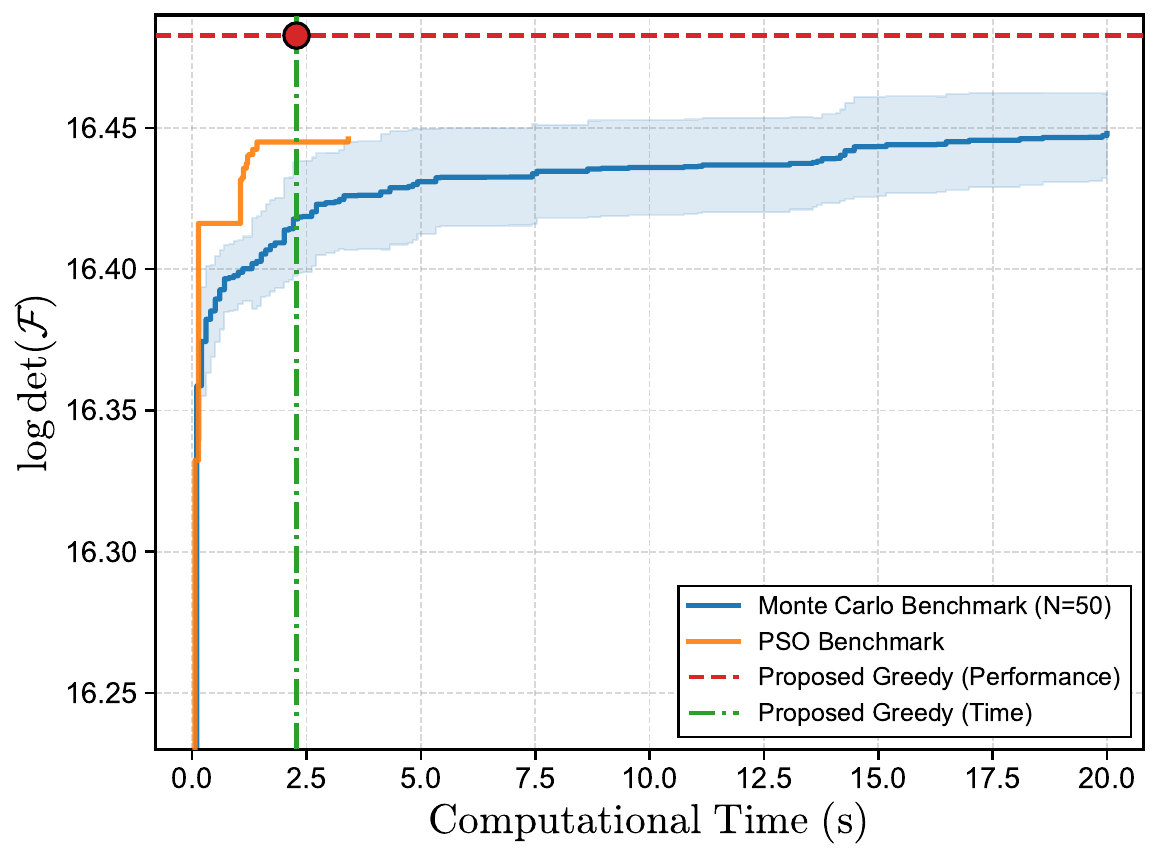} 
     \caption{Computational Efficiency Comparison: Proposed Greedy Algorithm vs. Monte Carlo and PSO Benchmarks} 
\label{Q1_time}
\end{figure}

Furthermore, we benchmark Alg.~\ref{Alg1} against a Monte Carlo (MC) search and Particle Swarm Optimization (PSO). Here, PSO is initialized with 30 particles over 50 iterations, and the MC search is configured with 50 independent stochastic trials. 
As illustrated in Fig.~\ref{Q1_time}, our approach locates a highly precise solution ($\log \det(\mathcal{F}) = 16.48$) in just $2.3$ seconds. Conversely, PSO prematurely converges to a suboptimal local maximum ($\approx 16.44$) after $3.5$ seconds, and the MC search plateaus at a comparable sub-optimal bound even after $20.0$ seconds. 
The Alg.~\ref{Alg1} also successfully satisfies the theoretical $(1-1/e)$-approximation bound ($16.482 > 12.36$) compared to the global optimum ($19.884$) obtained via exhaustive search. 
Our proposed algorithm bypasses both combinatorial explosion and local optima traps, making it highly suitable for real-time SWaP-constrained UAV deployments.

\begin{figure}
\centering
     \includegraphics[width=.35\textwidth]{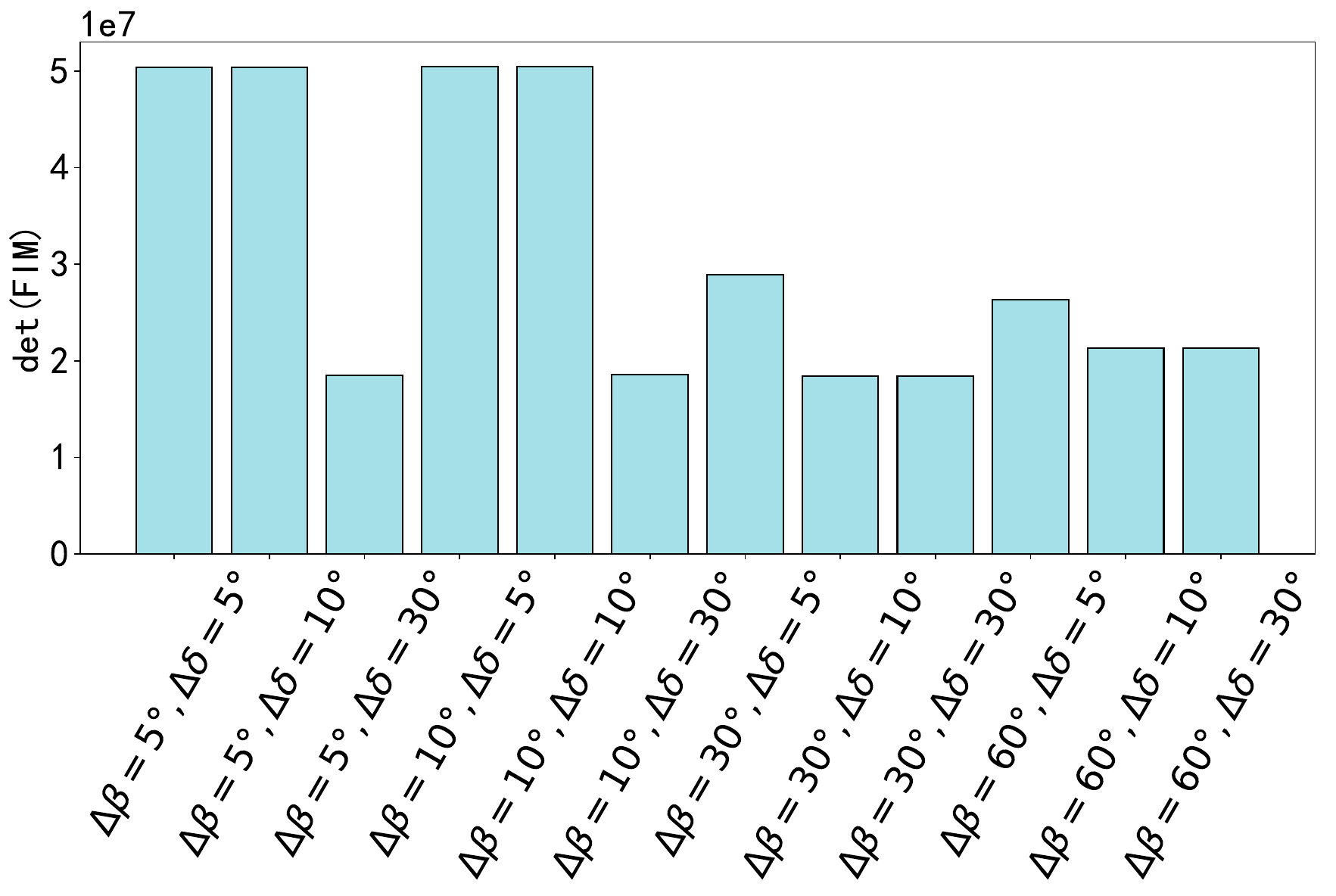} 
     \caption{Effect of Angular Discretization on Perception Performance} 
\label{sensitivity}
\end{figure}

To analyze the sensitivity to discretization, we conducted comparative experiments with varying angular resolutions. 
As shown in Fig.~\ref{sensitivity}, the coarse discretization (e.g., $\Delta\beta=60^\circ,\Delta\delta=30^\circ$) fails to identify optimal allocations, resulting in degraded perception accuracy. The moderate discretization (e.g., $\Delta\beta=10^\circ,\Delta\delta=10^\circ$) achieves an optimal balance between computational efficiency and solution quality. Notably, elevation angle discretization shows greater influence on accuracy than azimuth discretization, providing practical guidance for real-world implementations.

\begin{figure}
    \centering
    \includegraphics[width=0.7\columnwidth]{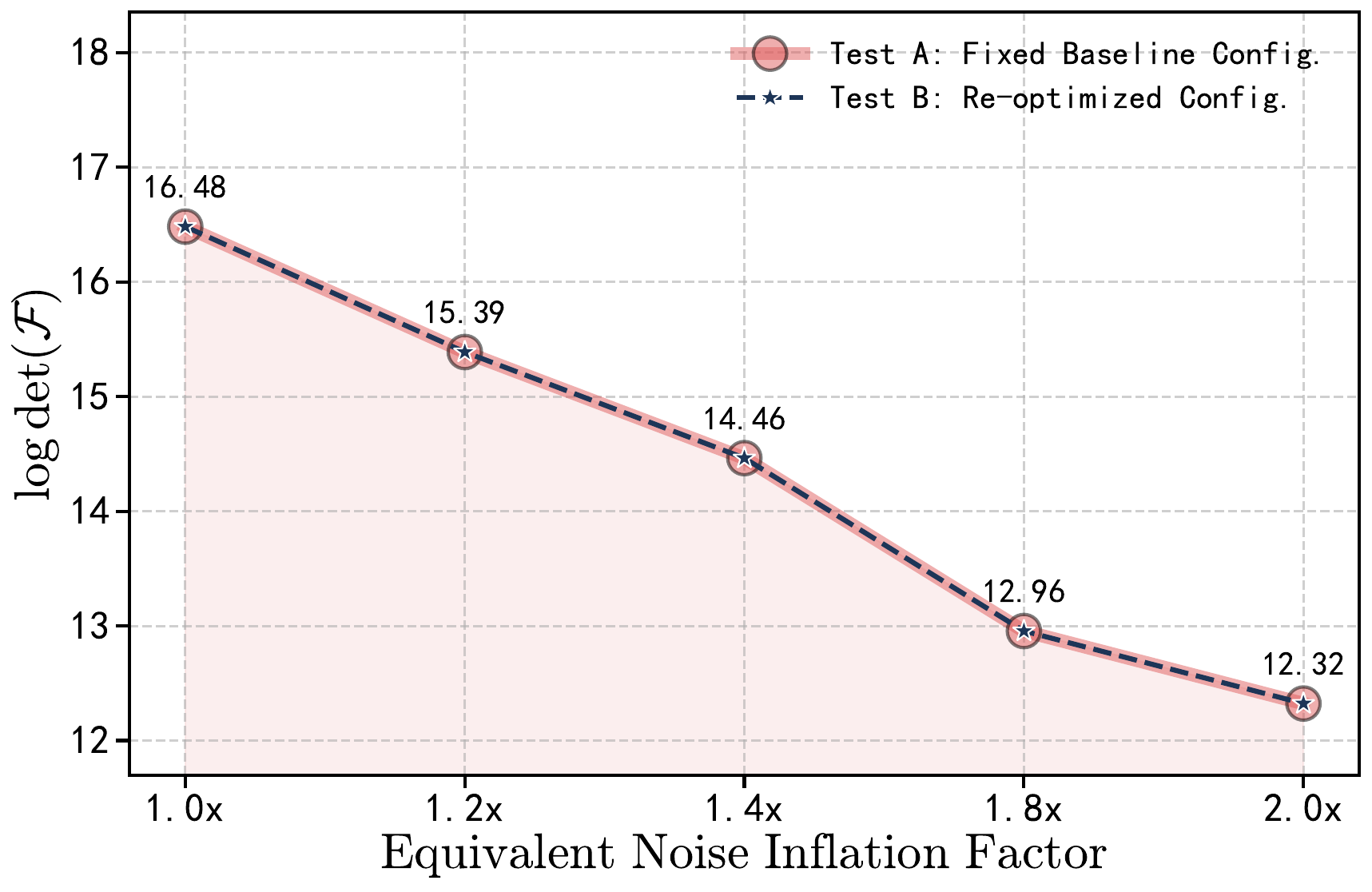}
    \caption{Dual-track sensitivity of $\log\det(\mathcal{F})$ under equivalent noise inflation from $1.0\times$ to $2.0\times$, for the fixed-baseline (Test A) and re-optimized (Test B) configurations.}
    \label{sensitivity2}
\end{figure}

In practical, minor timing misalignments and extrinsic calibration offsets are characterized as zero-mean additive perturbations that inflate the baseline measurement covariance matrices ($\mathbf{Q}_{\text{cam}}$ and $\mathbf{Q}_{\text{lidar}}$). To evaluate this impact, an equivalent noise inflation factor is applied to scale the covariance matrices from a 1.0x ideal baseline to a 2.0x extreme uncertainty condition.
Test A (Fixed Baseline Config.) where the optimal formation is derived at the 1.0x baseline, and Test B (Re-optimized Config.) where the algorithm re-computes the optimal formation under the inflated noise. Fig.~\ref{sensitivity2} gives the cumulative information gain that is quantified by $\log \det(\mathcal{F})$. This exhibits a predictable degradation (from 16.48 down to 12.32) as the equivalent noise intensifies.


\textcolor{black}{Remarkably, the performance degradation curves from both tests coincide. The dynamic re-optimization in Test B confirms that the optimal sensor allocation remains strictly invariant (comprising 2 LiDARs and 4 cameras) across the entire noise spectrum. 
The performance equivalence between the static and re-optimized topologies indicates that the original optimal formation retains a robust information geometry within the evaluated noise range. Consequently, upon real-world scenarios with minor calibration drift, the swarm can reliably maintain its nominal formation without initiating spatial reconfigurations.}


\begin{table}[htb]
\centering
\caption{UAV formation settings}
\label{table_uav_config1}
\begin{tabular}{ccccc}
\toprule
\textbf{UAV ID} & \textbf{Sensor} & \textbf{Azimuth ($^\circ$)} & \textbf{Elevation ($^\circ$)} & \textbf{Position ($x, y, z$)} \\
\midrule
1 & LiDAR  & 40.0  & 160.0 & $(-7.2, -6.0, 3.4)$ \\
2 & LiDAR  & 130.0 & 20.0  & $(-6.0, 7.2, 3.4)$  \\
3 & Camera & 0.0   & 160.0 & $(-9.4, -0.0, 3.4)$ \\
4 & Camera & 100.0 & 20.0  & $(-1.6, 9.3, 3.4)$  \\
5 & Camera & 50.0  & 160.0 & $(-6.0, -7.2, 3.4)$ \\
6 & Camera & 140.0 & 160.0 & $(7.2, -6.0, 3.4)$  \\
\bottomrule
\end{tabular}
\end{table}

\subsection{FOV-oriented Formation Performance}

\begin{figure*}[htbp]
\centering
\begin{subfigure}[b]{0.3\linewidth}
    \centering
    \includegraphics[width=\linewidth]{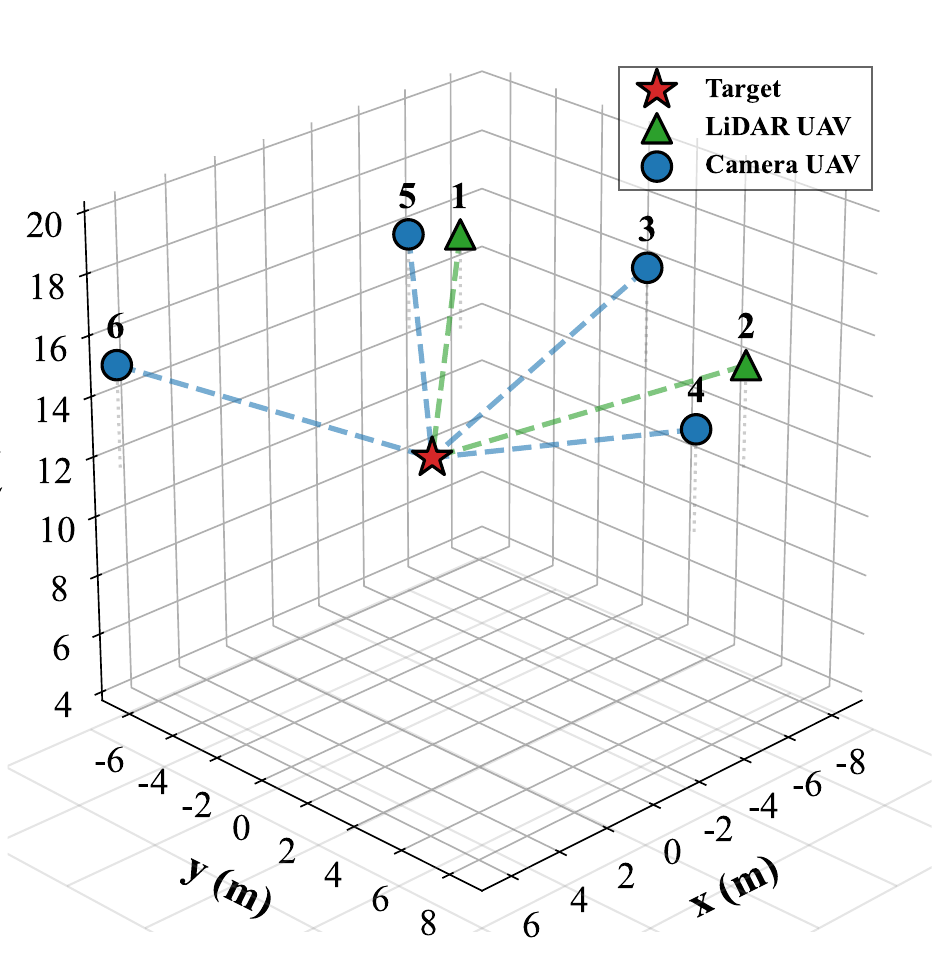}
    \caption{}
    \label{Q1_formation}
\end{subfigure}
\hfill
\begin{subfigure}[b]{0.3\linewidth}
    \centering
    \includegraphics[width=\linewidth]{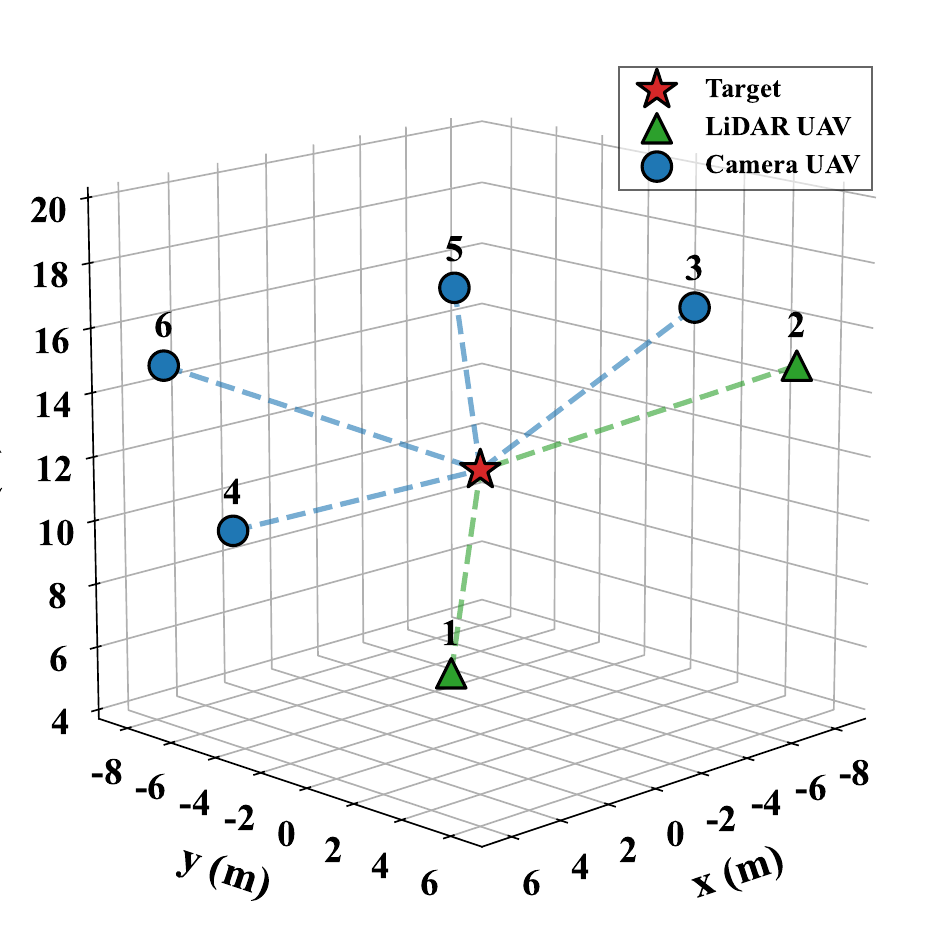}
    \caption{}
    \label{Q2_formation}
\end{subfigure}
\hfill
\begin{subfigure}[b]{0.3\linewidth}
    \centering
    \includegraphics[width=\linewidth]{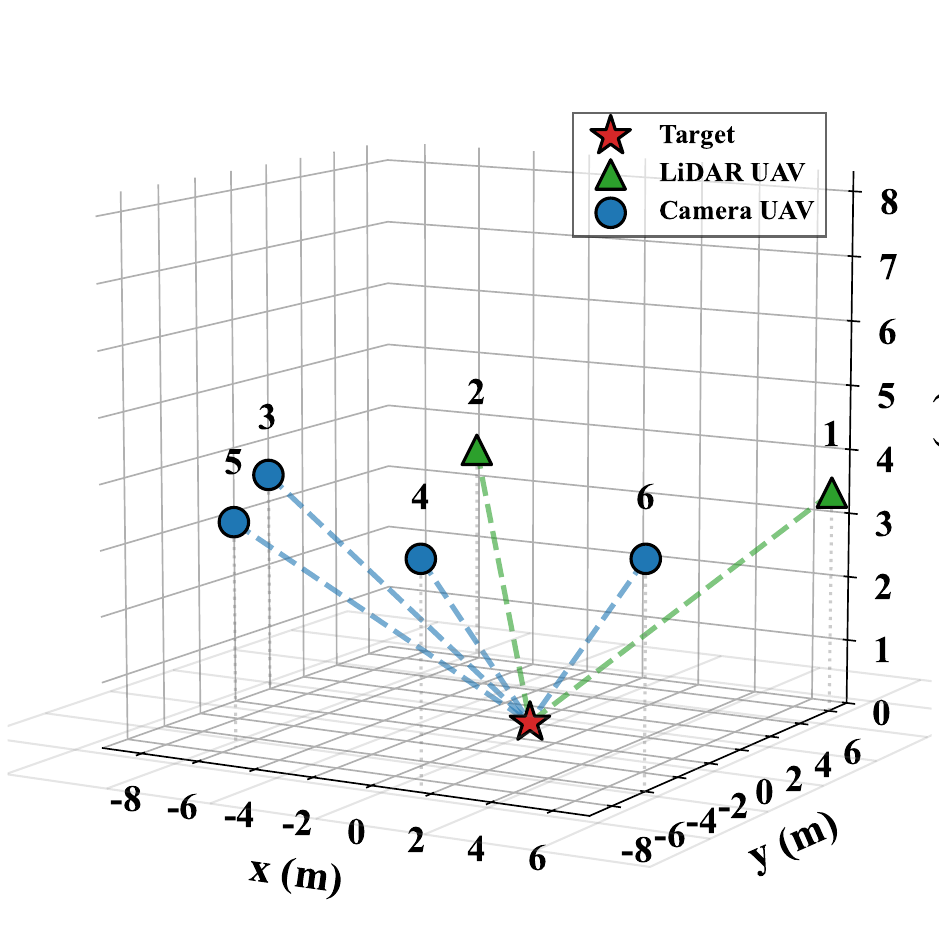}
    \caption{}
    \label{Q2_ground_formation}
\end{subfigure}
\caption{Heterogeneous UAV formation allocations (four camera-equipped UAVs as blue circles, two LiDAR-equipped UAVs as green triangles, target as red star): (a) initial formation produced by Alg.~\ref{Alg1}; (b) formation for an aerial target; and (c) formation for a ground target.}
\label{fig:formation_all}
\end{figure*}

Upon the UAV-sensor allocation derived in Sec. V-A, Alg.~\ref{Alg2} employs sector-constrained flipping operations (with $K=8$ azimuthal sectors) to explore the equivalent formation space while maintaining computational tractability.
Fig.~\ref{Q2_formation} shows the optimal formation for aerial target tracking. Alg.~\ref{Alg2} successfully transforms the initial UAV-sensor allocation into a FOV-oriented formation that eliminates coverage blind spots while preserving the FIM determinant. This spatial reconfiguration shows how equivalent geometric transitions can improve task effectiveness without compromising theoretical bounds.



For ground SAR scenarios, terrain constraints preclude UAV deployment below the target plane. To address this practical limitation, we implement a Z-axis reflection transition that repositions all UAVs to the feasible half-space while maintaining formation integrity. Although this constrained transition introduces a marginal reduction in sensing accuracy, the resulting allocation (Fig.~\ref{Q2_ground_formation}) achieves significantly enhanced FOV coverage and communication reliability compared to the initial allocation.




{\color{black}
Moreover, Tab.~\ref{tab:formation_performance_comparison} provides a performance comparison between our proposed formation and a benchmark uniform enclosure allocation. Since Alg.~\ref{Alg2} reconfigures the initial UAV-sensor allocation from Alg.~\ref{Alg1} through the equivalent transition, the FOV and SINR gains reported against Alg.~\ref{Alg1} reflect the effect of this transition alone, with the FIM determinant held fixed. 
The uniform enclosure (Bch.) places the same 6 UAVs (two LiDARs and four cameras) evenly around the target as an external geometric baseline, so that every reported percentage stems from formation geometry rather than resource endowment. For aerial targets, the optimal formation (Alg.~\ref{Alg2}) maintains the FIM log-determinant (Alg.~\ref{Alg1}) at 16.48 (superior to the uniform enclosure's 16.124), while simultaneously achieving a 25.0\% improvement in FOV coverage metric $\Gamma$ (increasing from 0.61 (Alg.~\ref{Alg1}) to 0.76 (Alg.~\ref{Alg2})). Most significantly, the average SINR increases by 104.2\% (from 6.97 (Alg.~\ref{Alg1}) to 14.23 dB (Alg.~\ref{Alg2})), with the minimum SINR rising from 1.78 dB (Alg.~\ref{Alg1}) to 11.51 dB (Alg.~\ref{Alg2}), thereby ensuring reliable communication across the entire formation control.}


In ground target scenarios, the Z-axis reflection transition yields a log-determinant of 16.414, marginally lower than the aerial target allocation but still exceeding the uniform enclosure's 16.241. The FOV coverage metric reaches 0.76, surpassing the uniform enclosure's 0.74, while the average SINR improves by 97.3\% relative to the initial state (reaching 13.78 dB). Although slightly lower than the uniform enclosure's 14.54 dB, this SINR level remains well within operational requirements for reliable data transmission.


\begin{figure*}[htbp]
\centering
\begin{subfigure}[b]{0.30\linewidth}
    \centering
    \includegraphics[width=\linewidth]{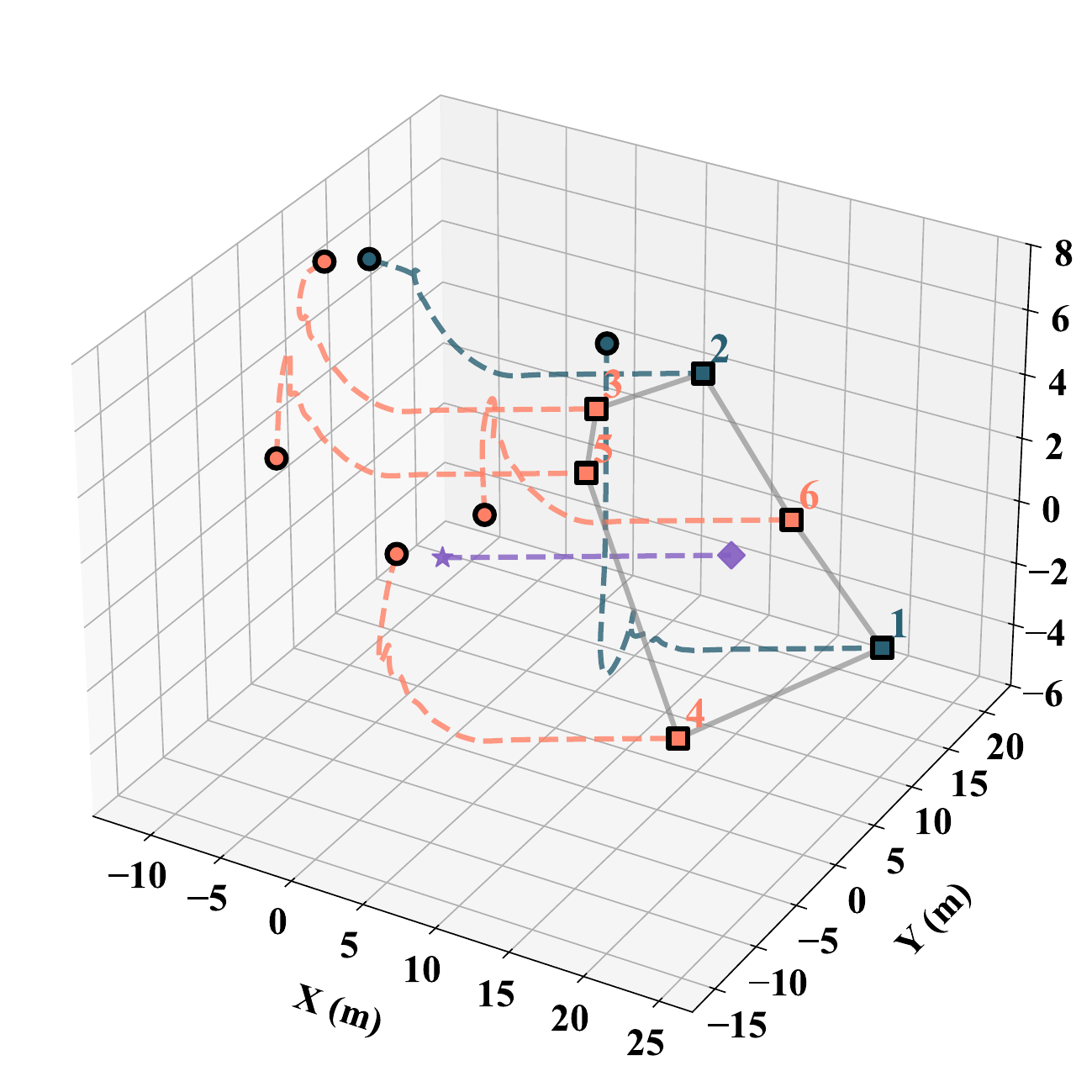}
    \caption{}
    \label{Pro_fomation}
\end{subfigure}%
\hfill
\begin{subfigure}[b]{0.30\linewidth}
    \centering
    \includegraphics[width=\linewidth]{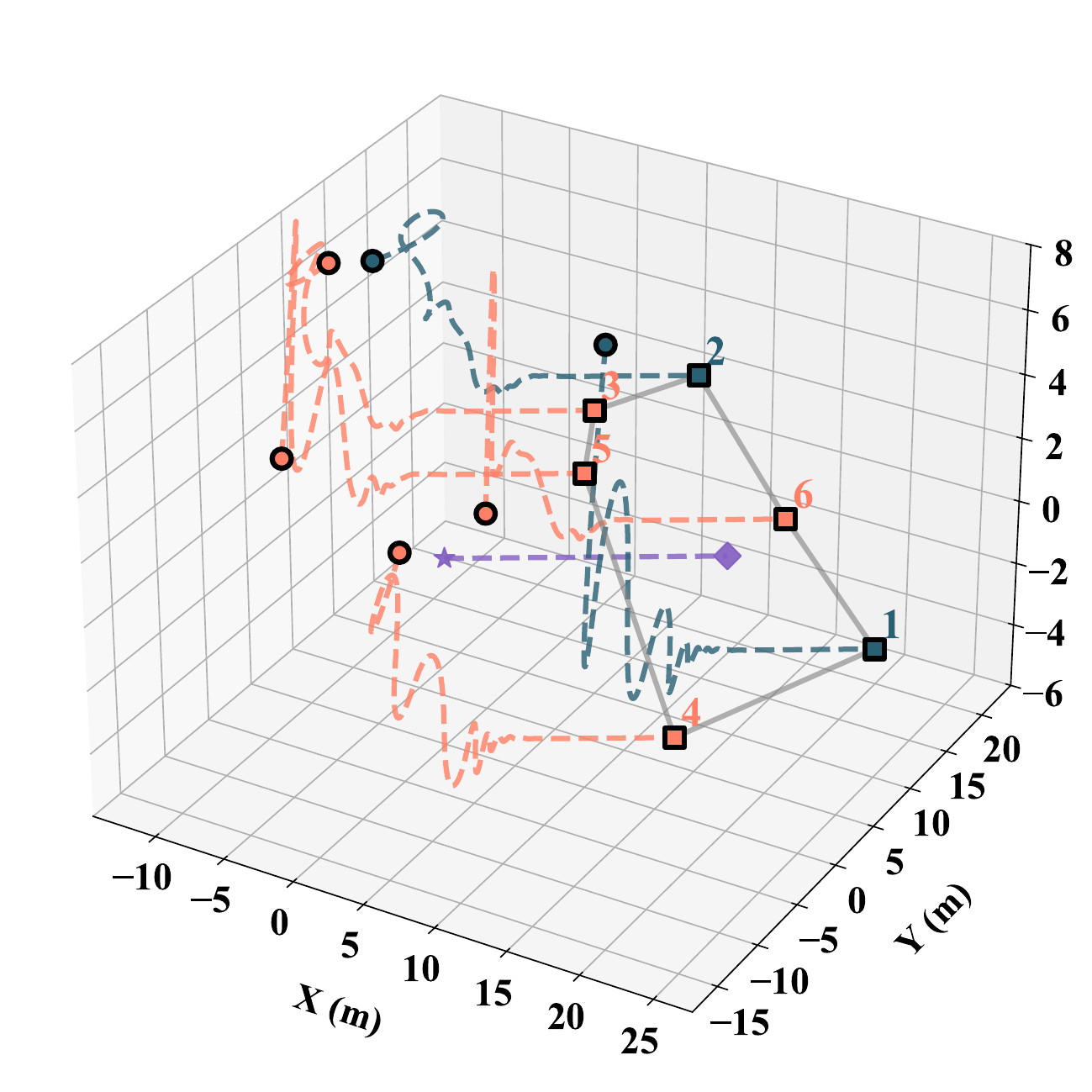}
    \caption{}
    \label{Quadratic_fomation}
\end{subfigure}%
\hfill
\begin{subfigure}[b]{0.30\linewidth}
    \centering
    \includegraphics[width=\linewidth]{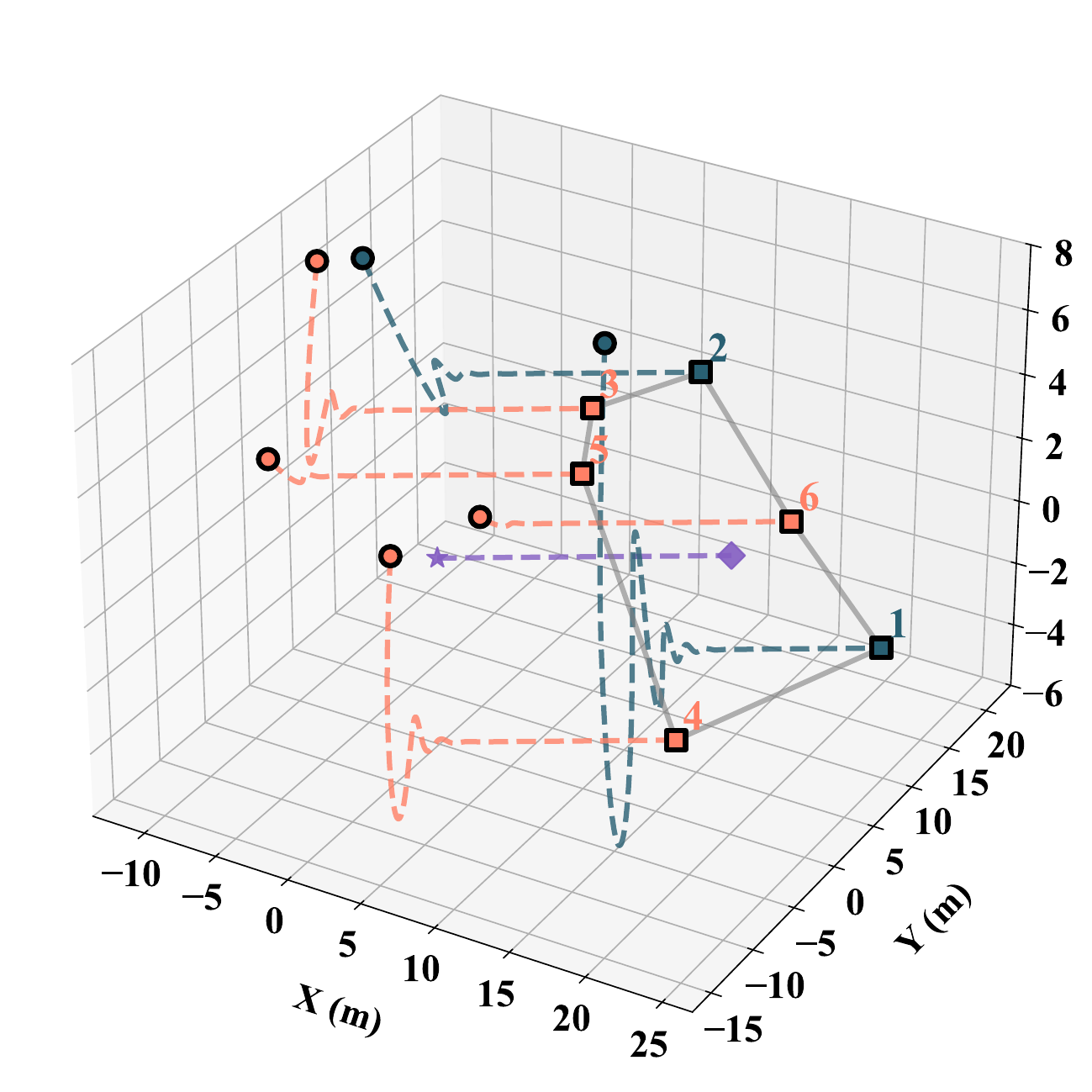}
    \caption{}
    \label{APF_fomation}
\end{subfigure}
\caption{3D trajectory evolution of the six UAVs during formation transition under (a) the proposed logarithmic-potential controller, (b) the quadratic-potential controller, and (c) the classical APF method. Purple star, dashed line, and diamond represent the target initial position, trajectory, and final position, respectively. Circles denote the UAV initial positions. Squares (labeled 1--6, matching Tab.~\ref{table_uav_config1}) denote the final positions. Gray lines represent the target formation. Blue and orange point represent the LiDAR UAVs and camera UAVs, respectively.}
\label{Q3_formation_comparison}
\end{figure*}


\begin{table}[htbp]
\centering
\caption{Performance of different formations}
\label{tab:formation_performance_comparison}
\setlength{\tabcolsep}{2.6pt} 
\begin{tabular}{lccccc}
\toprule
\textbf{Metric} & \textbf{Alg1} & \textbf{Alg2~(Air)} & \textbf{Alg2~(Gnd)} & \textbf{Bch.~(Gnd)} & \textbf{Bch.~(Air)} \\
\midrule
$\log\det(\mathcal{F})$ & 16.48 & 16.48 & 16.41 & 16.24 & 16.12 \\
$\Gamma$ & 0.61 & 0.76 & 0.76 & 0.74 & 0.74 \\
Avg. SINR           & 6.97   & 14.23  & 13.78  & 14.54  & 14.92  \\
Min. SINR          & 1.78   & 11.51  & 11.02  & 12.46  & 12.89  \\
\bottomrule
\end{tabular}

\vspace{5pt}
\footnotesize
\textbf{Note:} \textbf{Alg1} is the formation from Alg.~\ref{Alg1}; \textbf{Alg2} is the formation from Alg.~\ref{Alg2} (Air/Ground); \textbf{Bch.}: Uniform enclosure formation (Ground/Air).
\end{table}

\textcolor{black}{These results show that formation geometry optimization can simultaneously enhance multiple performance metrics, i.e., coverage completeness, communication reliability, and preservation of sensing accuracy. This supports our hypothesis that 3D spatial formation is an important rather than incidental component in multimodal swarm perception systems. }


\subsection{Flight Control Efficiency}

To validate the theoretical stability and energy efficiency of the proposed Lyapunov control in Sec. IV-C, we conducted simulations involving three distinct control strategies: (1) the proposed Lyapunov control with logarithmic potential function, (2) the conventional Lyapunov control with quadratic potential function, and (3) the classical Artificial Potential Field (APF) method. The simulation scenario includes a formation comprising 6 UAVs ($M=6$) with the optimal allocation derived from Alg.~\ref{Alg2}. 
To emulate a dynamic tracking of SAR operations, the formation tracks a maneuvering target trajectory $\mathcal{P}_{L}^{des}(t)$ with continuous velocity and directional variations. 
\textcolor{black}{Moreover, all three flight control schemes share the same target formation and the same target reference trajectory, so that the reported differences stem solely from the control scheme.}
Here, the quadratic control scheme is:
\begin{equation}
\mathbf{u}_i =
\begin{cases}
- k_1 \displaystyle\sum_{j\in\mathcal{M}_i}
(\mathbf{e}_{ij} )
- k_2 \mathbf{v}_i
- k_p(\mathcal{P}_i - \mathcal{P}_i^{des}),
& i= L, \\[8pt]
- k_1 \displaystyle\sum_{j\in\mathcal{M}_i}
(\mathbf{e}_{ij})
- k_2 \mathbf{v}_i,
& i\neq L,
\end{cases}
\label{eq:quad_control}
\end{equation}

\noindent where $L$ is the indicator of the leader UAV. 
In contrast, the APF method applies an attractive force guiding the UAV toward the desired position and a repulsive force for collision avoidance. The total APF force input is:
\begin{equation}
\mathbf{u}_i = - k_a (\mathcal{P}_i - \mathcal{P}_i^{*}) + \sum_{j\in \mathcal{M}_i} k_r\!\left(\frac{1}{\|\mathcal{P}_{ij}\|} - \frac{1}{d_0}\right) \frac{\mathcal{P}_{ij}}{\|\mathcal{P}_{ij}\|^3} - k_2 \mathbf{v}_i,
\label{eq:apf_control}
\end{equation}
\noindent where $\mathcal{P}_{ij}=\mathcal{P}_i - \mathcal{P}_j$ and $d_0$ is the distance of safe separation. The middle repulsive term is only active when $\|\mathcal{P}_{ij}\| < d_0$ and evaluates to zero when $\|\mathcal{P}_{ij}\| \ge d_0$. Furthermore, $k_a$ and $k_r$ are the attractive and repulsive gains. $k_2$ is the velocity damping gain to suppress oscillations.




\begin{figure*}[htbp]
\centering
\begin{subfigure}[b]{0.30\linewidth}
    \centering
    \includegraphics[width=\linewidth]{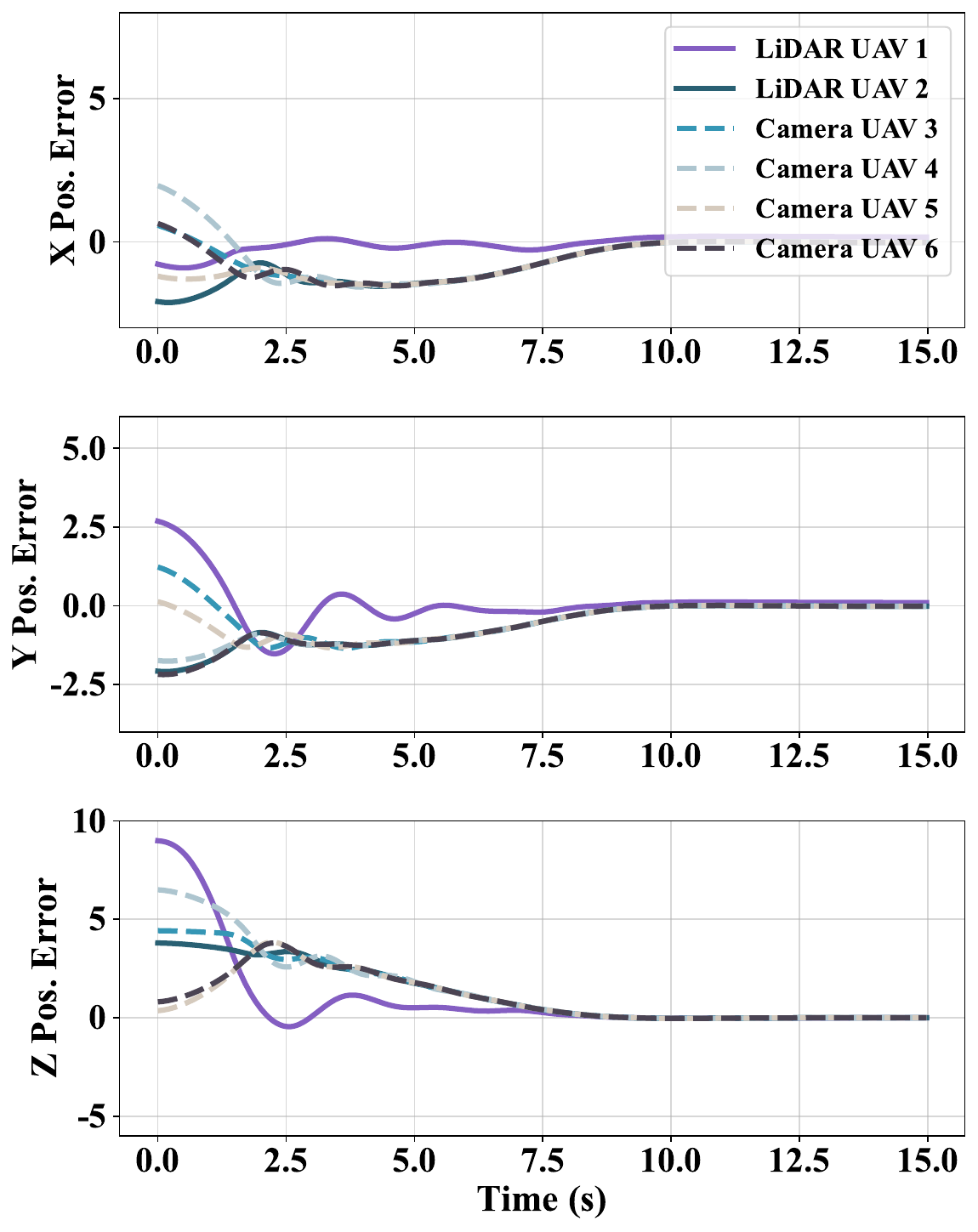}
    \caption{}
\end{subfigure}%
\hfill
\begin{subfigure}[b]{0.30\linewidth}
    \centering
    \includegraphics[width=\linewidth]{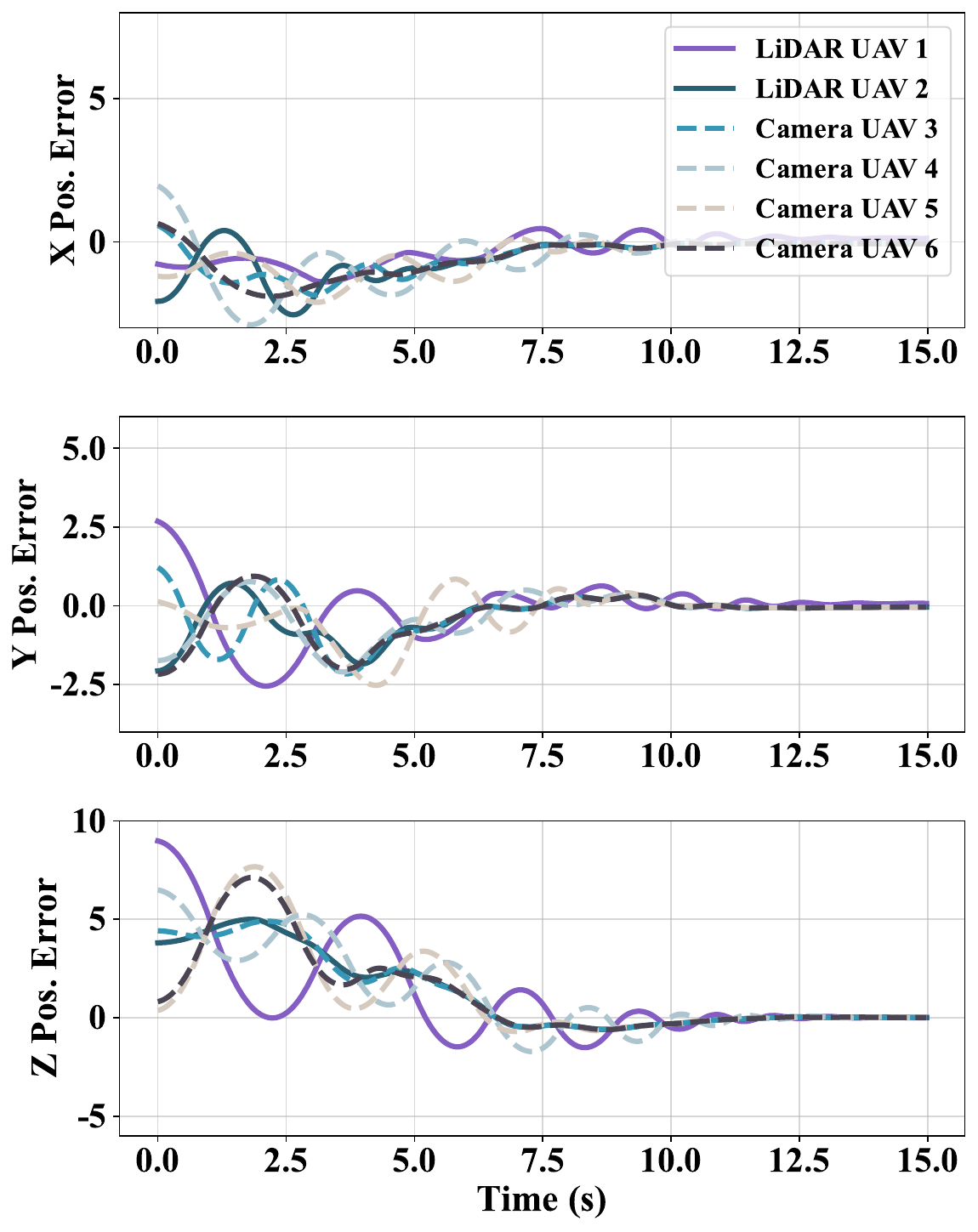}
    \caption{}
\end{subfigure}%
\hfill
\begin{subfigure}[b]{0.30\linewidth}
    \centering
    \includegraphics[width=\linewidth]{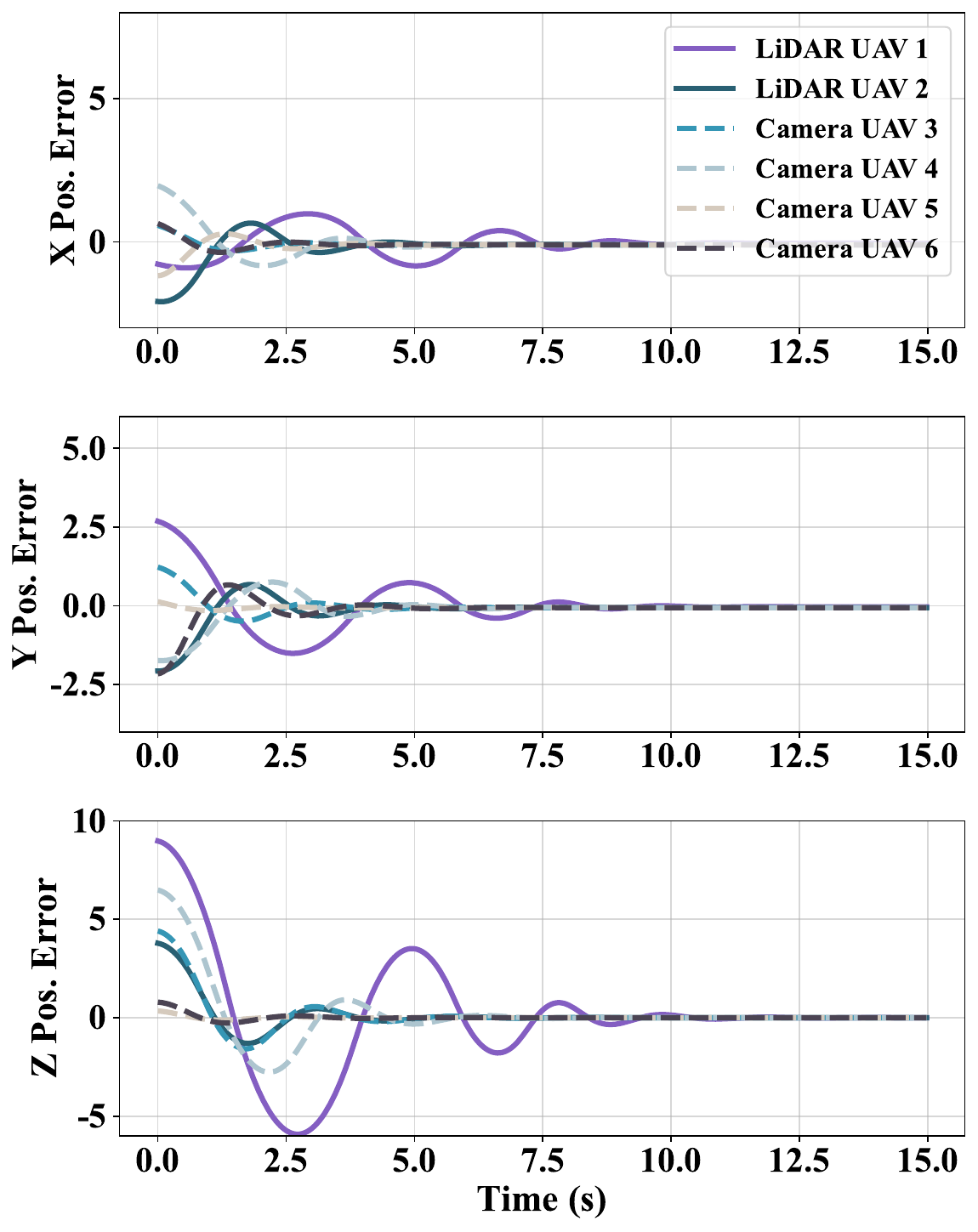}
    \caption{}
\end{subfigure}
\caption{3D position error evolution ($X$, $Y$, $Z$) of the six UAVs during formation transition under (a) the proposed logarithmic-potential controller, (b) the quadratic-potential controller, and (c) the classical APF method. Solid lines: LiDAR UAVs; dashed lines: camera UAVs.}
\label{Q3_position_comparison}
\end{figure*}

\begin{table}[htb]
\centering
\caption{Performance of different flight control}
\label{tab:performance_comparison}
\begin{tabular}{cccc}
\toprule
\textbf{Metric} & \textbf{Logarithmic} & \textbf{Quadratic} & \textbf{Classical APF} \\
\midrule
Avg. Distance (m)         & 15.62  & 29.60 & 19.40  \\
Avg. Velocity Err.       & 0.678  & 1.736 & 0.883  \\
Max. Velocity Err.       & 7.666  & 8.907 & 10.865 \\
Avg. Final Pos. Err. (m) & 0.056  & 0.085 & 0.107  \\
 $E_{\text{total}}$ ($\text{m}^2/\text{s}^3$) & 437.7 & 2858.8 & 1043.5 \\
\bottomrule
\end{tabular}
\end{table}

Tab.~\ref{tab:performance_comparison} summarizes the performance metrics during the formation transition. To evaluate energy efficiency, the total control energy $E_{\text{total}}$ over the transition period $T$ is given as:
\begin{equation}E_{\text{total}} = \sum_{i=1}^{M} \int_{0}^{T} |\mathbf{u}_i(t)|^2  dt,
\label{eq:energy_integral}
\end{equation}
\textcolor{black}{in which, minimizing $E_{\text{total}}$ involves a dual-factor optimization: suppressing the instantaneous control magnitude $\|\mathbf{u}_i(t)\|^2$ and reducing the total time $T$. The proposed logarithmic control scheme achieves both: its bounded control authority yields a smooth, near-constant velocity profile that restricts $\|\mathbf{u}_i(t)\|^2$, while the shortest travel trajectory proportionally reduces $T$.}

\textcolor{black}{The empirical results substantiate this dual-factor efficiency. Regarding $T$, the proposed method achieves the shortest average travel distance of 15.62 m, a 47.2\% reduction over the quadratic controller (29.60 m) and 19.5\% over the classical APF (19.40 m). Regarding the control magnitude, it attains an average velocity error of 0.678 m/s and a maximum of 7.666 m/s, lower than the quadratic (1.736 and 8.907 m/s) and APF (0.883 and 10.865 m/s) methods, alongside the smallest final position error of 0.056 m. Combining both factors, the control energy $E_{\text{total}}$ evaluated from the recorded inputs is only $437.7~\text{m}^2/\text{s}^3$ for the proposed method---an $84.7\%$ and $58.1\%$ reduction over the quadratic and APF baselines. It confirms that the logarithmic control scheme lowers energy not only by shortening the trajectory but also by bounding the instantaneous control magnitude $\|\mathbf{u}_i(t)\|^2$.}


Fig.~\ref{Q3_formation_comparison} shows the 3D trajectory evolution with various control schemes. The proposed logarithmic scheme (Fig.~\ref{Pro_fomation}) exhibits smooth convergence without oscillatory behavior. In contrast, both the quadratic (Fig.~\ref{Quadratic_fomation}) and APF scheme (Fig.~\ref{APF_fomation}) have pronounced oscillations during the transient phase. These oscillations necessitate additional corrective maneuvers that increase energy consumption and flight instability.


The position error dynamics in Fig.~\ref{Q3_position_comparison} provide further evidence of the proposed (logarithmic) control scheme's advantages. The proposed scheme achieves monotonic error reduction across all three spatial dimensions with minimal fluctuations, whereas both comparative methods exhibit significant oscillatory behavior before reaching steady state. While maintaining comparable convergence rates to the quadratic scheme, our approach eliminates the high-frequency oscillations that characterize conventional methods during the transient phase. 
This smooth error evolution directly contributes to reduced control effort, as evidenced by the lower maximum velocity error, and minimizes energy-intensive maneuvers.


\begin{figure}[t]
\centering
\includegraphics[width=0.48\linewidth]{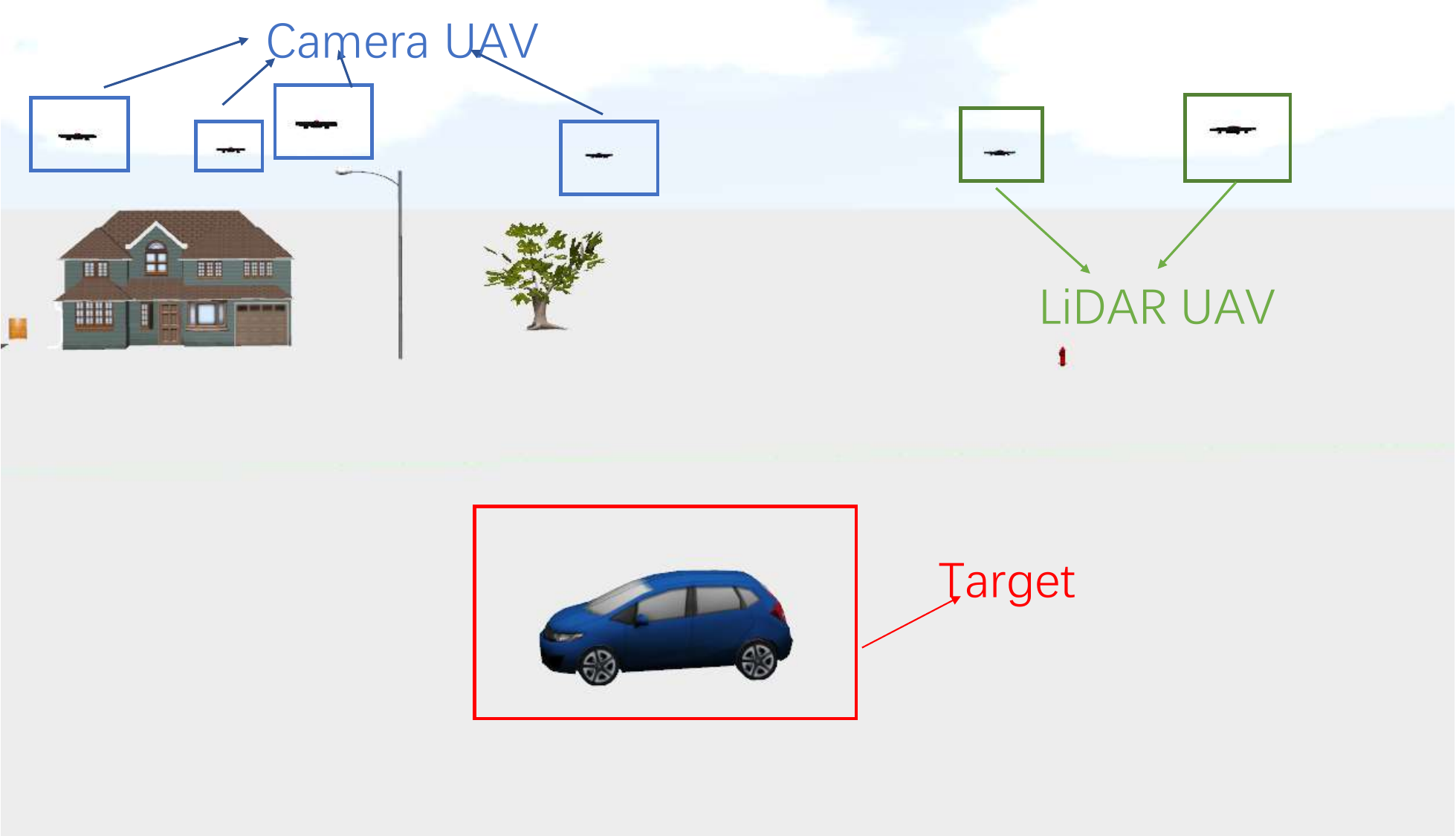}
\hfill
\includegraphics[width=0.48\linewidth]{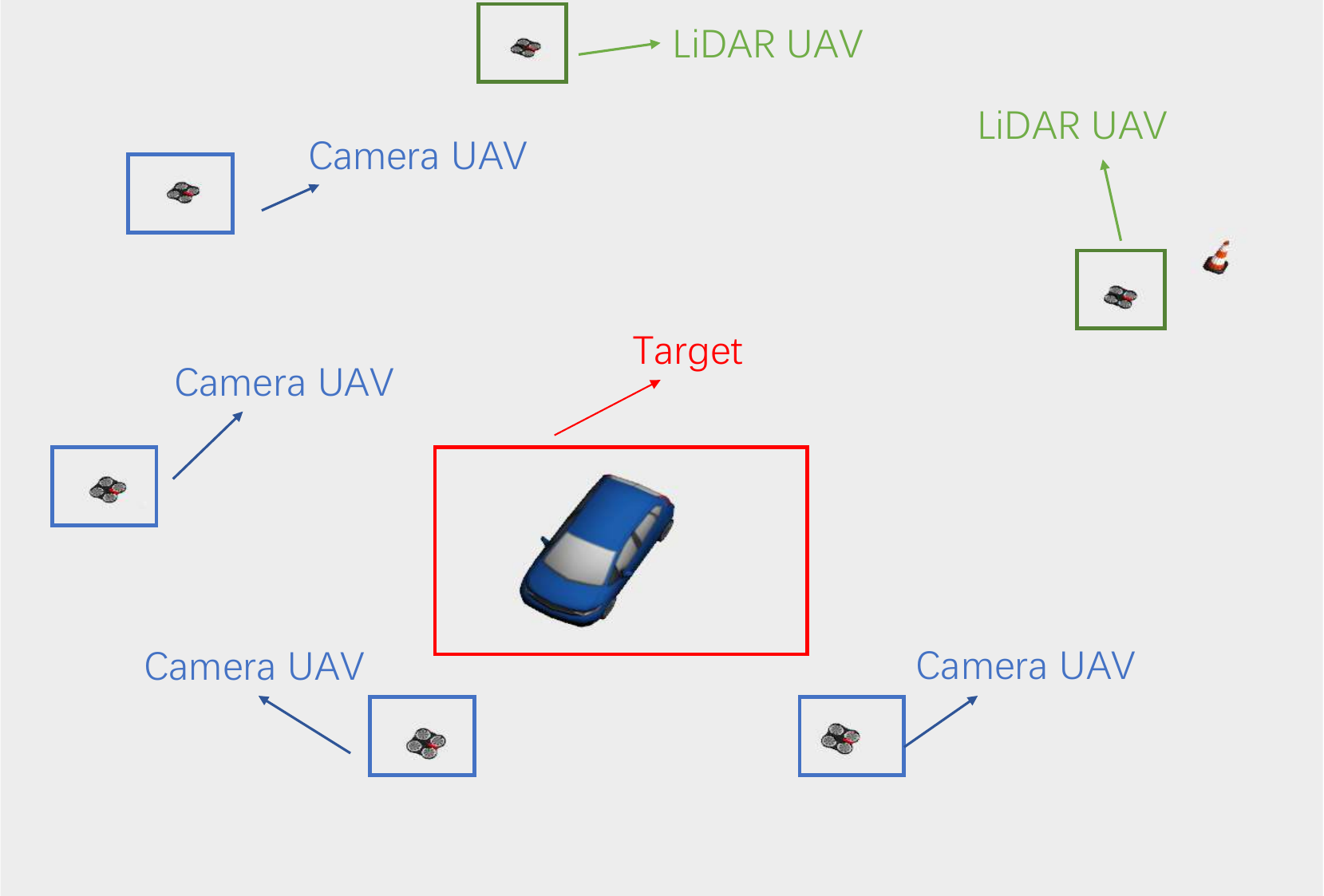}
\\[2pt]
\makebox[0.48\linewidth]{\small (a) Perspective view}
\hfill
\makebox[0.48\linewidth]{\small (b) Top-down view}
\caption{Gazebo settings: (a) perspective view; (b) top-down view of the optimal formation (two LiDARs and four cameras).}
\label{fig:gazebo_scene}
\end{figure}

\begin{table}[t]
\centering
\caption{Runtime of three modules}
\label{tab:runtime}
\begin{tabular}{llr}
\toprule
\textbf{Module} & \textbf{Setting} & \textbf{Runtime} \\
\midrule
UAV-sensor allocation & $|V|=1368$ & 2.30 s \\
Formation optimization & $M=6$ & 1.13 s \\
Flight-control update & $M=6$, per step & 1.11 ms \\
\bottomrule
\end{tabular}
\end{table}

\textcolor{black}{Finally, we assess the real-time feasibility of the three modules. As reported in Tab.~\ref{tab:runtime}, the one-shot allocation and formation-optimization steps complete within a few seconds, while the continuously running flight-control update takes only $1.11$~ms per step, far exceeding the $50$--$100$~Hz rate of practical controllers. Since the flight-control scheme requires only lightweight vector operations with a small memory footprint, it suits resource-constrained embedded controllers, whereas the one-shot steps can be executed pre-flight or on edge servers.}

\subsection{Task-Level Validation via Gazebo}

{\color{black}While the preceding evaluations validate the proposed framework through the information-optimal metric $\log\det(\mathcal{F})$, we further conduct a high-fidelity, task-level validation in the Gazebo simulation environment to bridge the gap between the geometric optimality and the cooperative perception performance. }
Typically, Gazebo is a widely adopted robotics simulator that provides a high-fidelity physics engine and photorealistic sensor rendering, enabling realistic emulation of UAV dynamics, camera imaging, and LiDAR point-cloud acquisition. 
It thus offers a faithful testbed for evaluating multi-UAV cooperative perception prior to costly physical deployment. 

To this end, we deploy the information-optimal formation derived from our framework into a photorealistic Gazebo world, where the 6 UAVs cooperatively perceive a ground target. Fig.~\ref{fig:gazebo_scene} illustrates the constructed scenario from two complementary viewpoints: a perspective view depicting the realistic sensing environment (Fig.~\ref{fig:gazebo_scene}(a)), and a top-down view showing the optimal formation geometry around the target (Fig.~\ref{fig:gazebo_scene}(b)). The multi-view fusion is performed using a distributed multi-view learning backbone~\cite{10233715}, where each UAV provides a single modality according to its onboard sensor: the two LiDAR UAVs encode their point clouds with PointNet, while the four camera UAVs encode their images with DenseNet121. The resulting 6 modality embeddings are then aggregated by a graph convolutional integrator to produce the final output.

\textcolor{black}{To reflect adverse real-world conditions, we inject additive Gaussian noise into the raw images and point clouds produced by the Gazebo sensors, thereby corrupting the inputs to the cooperative perception. }
Note that this data-level perturbation is distinct from the measurement covariance $\mathbf{Q}$ in Tab.~\ref{Simulation_parameters}: the latter characterizes the per-measurement statistical precision used in the FIM-based formation selection, whereas the former is applied to the raw sensor data to stress-test the end-to-end perception pipeline.

Hereafter, we evaluate the system under two test scenarios with progressively higher noise intensities. 
In Scenario~1 (S1), the image noise has mean $0.04$ and standard deviation $0.02$, while the point cloud noise has mean $0.02$ and standard deviation $0.05$. 
In scenario~2 (S2), both noise sources are further intensified, with the image noise mean raised to $0.06$ and the point-cloud noise mean raised to $0.04$, while the standard deviations remain at $0.02$ and $0.05$, respectively. 

Upon these settings, we benchmark the perception accuracy of the proposed optimal formation against 100 hemispherically sampled formations drawn from the same discrete candidate set ${V}$ defined in Sec.~IV-A, each consisting of 6 UAVs drawn under the identical heterogeneous sensor budget (two LiDARs and four cameras). 
For a fair comparison, the optimal formation is evaluated under the same number of independent trials with identical noise realizations, so that any performance difference is attributable solely to the formation geometry rather than to stochastic noise or sensor-budget variation. 
As summarized in Tab.~\ref{tab:gazebo}, Scenario~1 shows that the optimal formation attains an accuracy of $94.4\%$, surpassing the average of the hemispherical formations ($82.0\%$) by $12.4$ percentage points (a relative improvement of approximately $15.2\%$). In Scenario~2, the optimal formation still reaches $73.3\%$, outperforming the hemispherical formations ($61.0\%$) by $12.3$ percentage points (a relative improvement of approximately $20.2\%$). 
These task-level results confirm that the information-optimal geometry captures more discriminative multimodal features, yielding higher perception accuracy under noise suppression, with the benefit amplifying as observation quality deteriorates.

\begin{table}[t]
\caption{Perception accuracy in Gazebo.}
\label{tab:gazebo}
\centering
\renewcommand{\arraystretch}{1.3}
\setlength{\tabcolsep}{4pt}
\begin{tabular}{lcccccc}
\toprule
 & \multicolumn{2}{c}{\textbf{Image Noise}} & \multicolumn{2}{c}{\textbf{Point-Cloud Noise}} & \multirow{2}{*}{\textbf{Hemispherical}} & \multirow{2}{*}{\textbf{Optimal}} \\
\cmidrule(lr){2-3} \cmidrule(lr){4-5}
 & mean & std & mean & std & & \\
\midrule
S1 & 0.04 & 0.02 & 0.02 & 0.05 & 82.0\% & \textbf{94.4\%} \\
S2 & 0.06 & 0.02 & 0.04 & 0.05 & 61.0\% & \textbf{73.3\%} \\
\bottomrule
\end{tabular}
\end{table}

\section{Conclusion}


\textcolor{black}{This paper has shown that 3D formation geometry is an important rather than incidental design variable in multimodal UAV swarm cooperative perception during the active target tracking phase. By developing an information-optimal framework grounded in FIM determinant maximization, we jointly optimized heterogeneous sensor allocation, spatial allocation, and flight trajectory control. The framework integrates a submodular greedy algorithm with a $(1-1/e)$-approximation guarantee, an equivalent formation transition that enhances FOV coverage by 25.0\% and communication SINR by 104.2\% without sacrificing sensing accuracy. Additionally, a Lyapunov-stable logarithmic-potential controller that reduces control energy by 84.7\% over a quadratic-potential baseline. 
Upon noise suppression scenarios, the proposed formation improves perception accuracy by approximately 15.2\% and 20.2\% over 100 hemispherical formations from the same candidate set. 
While these evaluations validate the framework, high-fidelity simulation is not a substitute for physical deployment: real flight turbulence, on-board calibration, battery limits, and outdoor communication instability remain unmodeled. 
Future research will focus on hardware-in-the-loop flight experiments and adaptive formation reconfiguration under dynamic environmental occlusions.
}

\appendices
\footnotesize
\bibliography{biblio}
\end{document}